\documentclass[12pt]{article}
\usepackage[margin=.75in]{geometry}
\usepackage{amsmath, amssymb, amsthm}
\usepackage{mathtools}
\usepackage{graphicx}
\usepackage{array}
\usepackage{amsfonts}

\pdfoutput=1
\usepackage{putex}
\usepackage[vcentermath]{youngtab}
\usepackage{lscape}
\usepackage{cite}
\usepackage{graphicx}
\usepackage{epstopdf}
\usepackage{enumerate}
\usepackage{tensor}
\usepackage{slashed}
\usepackage{amsmath}
\usepackage{amssymb}
\usepackage{mathrsfs}
\usepackage{lgrind}
\usepackage{amsmath,amssymb,braket}
\usepackage{bbm}
\usepackage{hyperref}
\usepackage{subcaption}
\numberwithin{equation}{section}

\newcommand {\be} {\begin {equation}}
\newcommand {\ee} {\end {equation}}

\newcommand {\bes} {\begin {equation*}}
\newcommand {\ees} {\end {equation*}}

\newcommand{\eps}{\epsilon}

\def\CH{{\cal H}}

\newcommand{\beq}{\begin{equation}}
\newcommand{\eeq}{\end{equation}}

\newcommand{\sign}{\mathrm{sign}}

\def\be{ \begin{equation} }
\def\ee{ \end{equation} }

\def \be {\beta}

\def \beq { \begin{equation}}
\def \eeq {\end{equation}}

\usepackage{indentfirst}
\begin{document}
\preprint{PUPT-2585}
	
\institution{PU}{Department of Physics, Princeton University, Princeton, NJ 08544}
\institution{PCTS}{Princeton Center for Theoretical Science, Princeton University, Princeton, NJ 08544}
	
\authors{Igor R. Klebanov\worksat{\PU,\PCTS}, Preethi N. Pallegar\worksat{\PU},  Fedor K. Popov\worksat{\PU} }
\title{Majorana Fermion Quantum Mechanics for Higher Rank Tensors}

\abstract{We study quantum mechanical models in which the dynamical degrees of freedom are real fermionic tensors of rank five and higher. They are the non-random counterparts of the Sachdev-Ye-Kitaev (SYK) models where the Hamiltonian couples six or more fermions. For the tensors of rank five, there is a unique $O(N)^5$ symmetric sixth-order Hamiltonian leading to a solvable large $N$ limit dominated by the melonic diagrams. We solve for the complete energy spectrum of this model when $N=2$ and deduce exact expressions for all the eigenvalues. The subset of states which are gauge invariant exhibit degeneracies related to the discrete symmetries of the gauged model. We also study quantum chaos properties of the tensor model and compare them with those of the $q=6$ SYK model. For $q>6$ 
there is a rapidly growing number of $O(N)^{q-1}$ invariant tensor interactions. We focus on those of them that are maximally single-trace --- their 
stranded diagrams stay connected when any set of $q-3$ colors is erased. 
We present a general discussion of why the tensor models with maximally single-trace interactions have large $N$ limits dominated by the melonic diagrams.
We solve the large $N$ Schwinger-Dyson equations for the higher rank Majorana tensor models and show that they match those of the corresponding SYK models exactly. We also study other gauge invariant operators present in the tensor models. 
}
	
\date{}
\maketitle
\tableofcontents
\section{Introduction and summary}

In recent literature, quantum mechanical models in which the dynamical degrees of freedom are fermionic tensors of rank $3$ and higher 
have attracted much attention, starting with the papers \cite{Witten:2016,Klebanov:2016xxf}. These theories can have interesting large $N$ limits where the dominant ``melonic" diagrams can be simply studied and summed \cite{Gurau:2009tw,Gurau:2011aq,Gurau:2011xq,Bonzom:2011zz,Tanasa:2011ur,Bonzom:2012hw,Carrozza:2015adg}
(for reviews, see  \cite{Gurau:2011xp,Tanasa:2015uhr,Delporte:2018iyf,Klebanov:2018fzb}). 
In the $O(N)^3$ symmetric quantum mechanical model for Majorana fermions \cite{Klebanov:2016xxf}, 
the unique non-trivial quartic term has the tetrahedral structure 
\begin{gather}
H_4 = \frac{g_t}{4} \psi^{abc}\psi^{ab'c'}\psi^{a'bc'}\psi^{a'b'c}\ , \label{q4Ham}
\end{gather} 
where each of the indices ranges from $1$ to $N$ and the repeated indices are summed over.\footnote{There are also three quartic terms of ``pillow" topology \cite{Carrozza:2015adg}; they are the quadratic Casimir operators of the three $SO(N)$ groups \cite{Bulycheva:2017ilt} and are, therefore,
determined by the group representation.  In the gauged model they vanish.}
In the large $N$ limit where $g N^{3/2}$ is held fixed, the surviving Feynman diagrams are melonic, and they can be summed using Schwinger-Dyson equations.
These diagrams are the same as in the Sachdev-Ye-Kitaev (SYK) model \cite{Sachdev:1992fk,Kitaev:2015,Sachdev:2015efa,Polchinski:2016xgd,Maldacena:2016hyu,Kitaev:2017awl}, where the quartic interactions contain a random four-index tensor. 
As a result, the large $N$ tensor and SYK models are closely related, although there are also some important differences \cite{Klebanov:2018fzb}. These differences
are manifest in the small $N$ exact diagonalizations of the Hamiltonians \cite{Krishnan:2016bvg,Krishnan:2017txw,Klebanov:2018nfp,Krishnan:2018hhu,Krishnan:2018jsp,Pakrouski:2018jcc}. 

Rank $q-1$ tensor models with $q>4$ have been the subject of several studies relevant to our paper \cite{Klebanov:2016xxf,Yoon:2017,Ferrari:2017jgw,Gubser:2018yec}.
A comprehensive study of various invariant interaction vertices for a single tensor of rank $q-1$  was carried out in \cite{Gubser:2018yec,Ferrari:2017jgw}. 
For $q \geq 8$ there is a very rapidly growing number of ``generalized tetrahedral" interaction vertices, i.e. those that satisfy the constraint that every pair of tensors has exactly one index contraction.\footnote{This is to be contrasted with the Gurau-Witten model \cite{Gurau:2009tw,Witten:2016} for $q$ flavors of rank $q-1$ Majorana fermion tensors, where 
the interaction is uniquely fixed by the
$O(N)^{q(q-1)/2}$ symmetry.} As pointed out in \cite{Yoon:2017}, their counting is a mathematical problem isomorphic to scheduling of the round-robin tournament.
 Following \cite{Ferrari:2017jgw} we mostly focus on the special subclass of such interactions which are
``maximally single-trace" --- their stranded diagrams stay connected if any set of $q-3$ colors is erased.
As we discuss in section \ref{melonic_dominance},
 this facilitates the combinatorial analysis of the Feynman diagrams in the large $N$ limit. It is conjectured that the 
maximally single-trace (MST) interaction vertices, which are known in mathematical literature as perfect $1$-factorizations, exist for any even $q>2$. They have been proven to exist when either $q-1$ or $q/2$ is prime \cite{midori88,colbourn2010crc}, as well as in some other cases, such as $q=16, 28, 36, 40, 50, 126, 170$, etc.

\newcommand{\T}{\mathcal{T}}

\begin{figure}
\centering
{\includegraphics[width=0.25\textwidth]{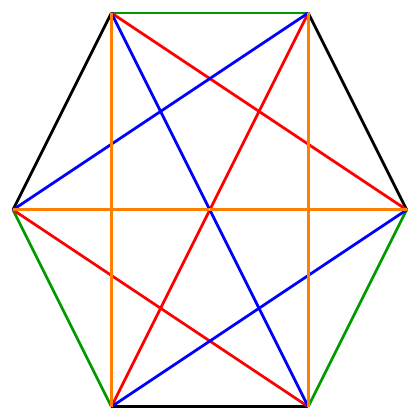}}
\caption{\label{fig:tikzsix} A graphical representation of the unique ``generalized tetrahedral" interaction for $q=6$, given in \eqref{Hamil}. Each line represents an index contraction, 
while different colors correspond to different groups. This interaction is maximally single-trace, since erasing any set of three colors leaves the diagram connected.}
\end{figure}

A part of our paper is devoted to a careful analysis of the Majorana tensor theory in $0+1$ dimension with rank-$5$ tensors as the dynamical degrees of freedom. The unique generalized tetrahedral interaction was written down in \cite{Klebanov:2016xxf}, and the hermitian Hamiltonian is
\begin{equation} \label{Hamil}
\begin{split}
H_6 = i \frac g2 \big (&
\psi^{a_1 b_1 c_1 d_1  e_1}\psi^{a_1 b_2 c_2 d_2 e_2} \psi^{a_2 b_2 c_3 d_3 e_1} 
\psi^{a_2 b_3 c_2 d_1 e_3}\psi^{a_3 b_3 c_1 d_3 e_2} \psi^{a_3 b_1 c_3 d_2 e_3}- \\ 
&  \psi^{a_3 b_1 c_3 d_2 e_3} \psi^{a_3 b_3 c_1 d_3 e_2} \psi^{a_2 b_3 c_2 d_1 e_3} \psi^{a_2 b_2 c_3 d_3 e_1} \psi^{a_1 b_2 c_2 d_2 e_2} \psi^{a_1 b_1 c_1 d_1  e_1}
\big )
\ .
\end{split}
\end{equation}
We can graphically depict this interaction by representing each fermion as a vertex of a graph, and each index contraction between pairs of fermions as an edge connecting two vertices (see fig. \ref{fig:tikzsix}). In the large $N$ limit, where $\lambda^2=g^2N^{10}$ is held fixed, the melonic diagrams dominate. 
The factor of $i$ is necessary to make $H_6$ real; it is a new feature compared to the rank-$3$ Hamiltonian (\ref{q4Ham}).
The Hamiltonian (\ref{Hamil}) has $SO(N)^5$ symmetry, as well as some discrete symmetries. Some aspects of this tensor model are similar to the $O(N)^3$ tensor model. The energy spectra in both models are symmetric under $E\rightarrow -E$, since an interchange of any two of the $O(N)$  groups sends $H \rightarrow -H$.  
However, there are also some differences: for example, in the $O(N)^5$ model the time-reversal is not a symmetry since it acts as $\T^{-1} H \T=-H$ due to the factor $i$ present in
the Hamiltonian (\ref{Hamil}). 

The $O(N)^5$ model also has some differences from the $q = 6$ SYK model. In particular, at small $N$ the structure of the spectra are rather different. This is due to the large number of continuous and discrete symmetries, which makes the tensor spectrum highly degenerate.  
The $q = 6$ SYK spectrum is compared with the corresponding Q-hermite polynomial, which is calculated in the double scaling limit, where $N_{\rm SYK}  \rightarrow \infty$, $q \rightarrow \infty$ with $q^2/N_{\rm SYK} $ held fixed \cite{Cotler:2016fpe,Berkooz:2018jqr}. 
We find very good agreement, which suggests that the $q = 6$ SYK model shares some spectral properties with the double scaled model.

The structure of the paper is as follows. 
In section \ref{classical} we discuss the structure of the Hamiltonian  \eqref{Hamil} and its symmetries and use them to explain some of the degeneracies that we observe in the singlet spectrum in section \ref{singlets}. In section \ref{finite_N} we numerically study the spectrum of the tensor model and the $q=6$ SYK model and investigate the differences between the spectral properties at finite $N$. In section \ref{higher_q} we discuss some properties of higher $q$ tensor models.
In section \ref{melonic_dominance} we present a simple diagrammatic argument for the melonic dominance for the maximally single-trace vertices. The Schwinger-Dyson equations of the $O(N)^5$ and $O(N)^7$ models are computed in section \ref{large_N}. We show the existence of the solution of these equations in the IR limit, and that it is invariant under conformal transformations. Additionally, we study the spectrum of the singlet bilinear and some of the non-singlet bilinears and show that they are identical to the SYK model.

\section{Hamiltonian and its symmetries}\label{classical}

The model contains a set of $N^5$ Majorana fermions $\psi^{abcde}$ with
the anti-commutation relations: 
\begin{align}
\{ \psi^{abcde}, \psi^{a'b'c'd'e'}\} = \delta^{aa'}\delta^{bb'}\delta^{cc'} \delta^{dd'}\delta^{ee'} \, . \label{comrel}
\end{align}
We will first work at the ``classical level", where we ignore the delta-function terms 
on the RHS of \eqref{comrel}  and treat the fermions as real grassman numbers.
Then the Hamiltonian we consider is
\begin{align}
H_{\rm class} = i g
\psi^{a_1 b_1 c_1 d_1  e_1}\psi^{a_1 b_2 c_2 d_2 e_2} \psi^{a_2 b_2 c_3 d_3 e_1} 
\psi^{a_2 b_3 c_2 d_1 e_3}\psi^{a_3 b_3 c_1 d_3 e_2} \psi^{a_3 b_1 c_3 d_2 e_3}
\  .
\label{classH}
\end{align}
This is the unique sextic term with $O(N)^5$ symmetry where any pair of fields have one index contraction \cite{Klebanov:2016xxf}. The factor $i$ is inserted so that the Hamiltonian is hermitian.
The correct quantum Hamiltonian (\ref{Hamil}) is $H_{6} = H_{\rm class}+H^\dagger_{\rm class}$. 

We can show that $H_{\rm class}$ changes sign under permutation of two $O(N)$ groups.
For example, when we permute $O(N)_c$ and $O(N)_d$, the fermions transform as
\begin{align}
\psi^{abcde} \leftrightarrow \psi^{a b  d c e}.
\end{align}
So, $H_{\rm class}\rightarrow H'_{\rm class}$ where
\begin{align}
H'_{\rm class}= i g
\psi^{a_1 b_1 c_1 d_1  e_1}\psi^{a_1 b_2 c_2 d_2 e_2} \psi^{a_2 b_2 c_3 d_3 e_1} 
\psi^{a_2 b_3 c_1 d_2 e_3}\psi^{a_3 b_3 c_3 d_1 e_2} \psi^{a_3 b_1 c_2 d_3 e_3}
\ .
\end{align}
Dropping the quantum delta-function terms in (\ref{comrel}), and bringing it to the form so that  the fields are read from right to left, we have
\begin{align}
H'_{\rm class}= - i g
\psi^{a_3 b_1 c_2 d_3 e_3} \psi^{a_3 b_3 c_3 d_1 e_2} \psi^{a_2 b_3 c_1 d_2 e_3}
\psi^{a_2 b_2 c_3 d_3 e_1} \psi^{a_1 b_2 c_2 d_2 e_2}
\psi^{a_1 b_1 c_1 d_1  e_1}  
\ .
\end{align}
We find that $H'_{\rm class}= -H_{\rm class}$; this can be seen explicitly by relabeling the indices 
\begingroup
\begin{equation}
\begin{split}
& a_1 \leftrightarrow a_3, \quad  e_1 \leftrightarrow e_3, \quad  b_2 \leftrightarrow b_3, \\
& c_2 \rightarrow c_1, \quad c_1 \rightarrow c_3, \quad c_3 \rightarrow c_2,  \\
& d_3 \rightarrow d_1, \quad d_1 \rightarrow d_2, \quad d_2 \rightarrow d_3 .
\end{split}
\end{equation}
\endgroup
We examine the behavior under the other $O(N)$ permutations and find that $H' = -H$ in all cases, so the symmetry group of the Hamiltonian includes the alternating group $A_5$. This is related to the fact that it is a maximally single-trace (MST) operator. We expect that the $A_{q-1}$ symmetry also holds for the MST Hamiltonians with higher even $q$.

When we use the quantum anti-commutation relations \eqref{comrel}, the Hamiltonian (\ref{classH}) is not hermitian.
Adding the hermitian conjugate, we find (\ref{Hamil}). It is then possible to check that under a permutation of two indices $H_6\rightarrow -H_6$, establishing the  $A_5$
symmetry at the quantum level.
In the second term of $H_6$ we may bring the variables back into the same position as in the first term. To do this we need to make 15 permutations, which give rise to 15 additional quartic terms. Indeed, we can add possible quartic terms to the quantum Hamiltonian (\ref{Hamil}), but as it is shown in the appendix  \ref{quartic}, they do not preserve the $A_5$ symmetry mentioned above. 
 The Hamiltonian \eqref{Hamil} can be also obtained via the path integral formulation of the model with real grassmanian variables, and by calculating the corresponding operator by Weyl ordering. Another way to see this is to notice that this is the only operator up to the sixth order in fermions that respects the $A_5$ symmetry. 

We may choose the representation where each $\psi^{abcde}$ is a hermitian matrix with real entries. Thus, in a given basis (\ref{Hamil}) is a hermitian matrix with imaginary entries; therefore, it is antisymmetric. 
This means that its eigenvalues are real and come in pairs $\pm E$. This implies that the spectrum has
symmetry under $E\rightarrow -E$, which is a desired property. The proof is the following: let us start with some real matrix, $H'$. From $H'$ we can construct a hermitian matrix, $H=i(H'-H'^T)$. All entries of this matrix are complex, $H=-H^*$, and by definition, $H^\dagger = H$, where $H^\dagger$ is the adjoint. We can therefore conclude that $H^T=-H$. We write the characteristic equation:
\begin{equation}
\det(H-\lambda I) = 0 \Rightarrow \det(H^T -\lambda I) = \det( H + \lambda I) = 0
\end{equation}
Thus we have shown that the energy spectrum of (\ref{Hamil}) is symmetric. Another way to see this is to consider the time reversal symmetry, which we discuss in the following section.

\subsection{Discrete symmetries}\label{sub_discrete}
As in \cite{Klebanov:2018nfp} we can introduce an operator that sends $H \rightarrow -H$. This is called the permutation operator, and it implements an $O(N)$ group pair swap. We can implement this operation by introducing the following operator
\begin{align}
P_{45} =  \prod_{a,b,c,d=e}\psi^{abcde}\prod_{a,b,c,d>e}\left(\frac{\psi^{abcde}+\psi^{abced}}{\sqrt{2}}\right), \quad P_{45}^\dagger \psi^{abcde} P_{45} = \psi^{abced},
\end{align}
which exchanges the last two indices of each fermion in the interaction. 

For convenience, it is better to work with Dirac fermions, which can be built in the following way
\begin{gather}
\psi^{abcd n} = \frac{\psi^{abcd(2n-1)}+i\psi^{abcd(2n)}}{\sqrt{2}},\quad \bar{\psi}^{abcdn}=\frac{\psi^{abcd(2n-1)}-i\psi^{abcd(2n)}}{\sqrt{2}},
\end{gather}
and they satisfy the usual commutation relations of the Dirac fermions.

We notice a symmetry under the exchange of $\psi^{abcd (2n)} \to -\psi^{abcd(2n)}$ in Hamiltonian (\ref{Hamil}). It corresponds to the charge conjugation symmetry, $C$, $\bar{\psi}^{abcd n} \leftrightarrow \psi^{abcdn}$. Under this exchange, each term gains a negative sign during normal ordering, and this results in preservation of the original Hamiltonian: $CHC^{-1} = H$. We can define the charge conjugation operator, 
\begin{align}
C = \prod_{abcdn}\psi^{abcd(2n-1)}.
\end{align}

In the case of the $q=4$ tensor model \cite{Klebanov:2018nfp} there is an anti-unitary time-reversal symmetry $T$, that acts in the following way
$$
T i T^{-1} = -i, \quad T \psi^{abcde} T^{-1} = \psi^{abcde}, \quad T H_{4} T^{-1} = H_{4}.
$$
In the case of the Hamiltonian \eqref{Hamil} this is not a symmetry of the theory. Indeed,
$$
T H_{6} T^{-1} = -H_{6},
$$
which shows that $T$ is not a symmetry of the theory. From this one can see that the eigenvectors come in the pairs $(\ket{E}, T\ket{E}$) with opposite energies. In the representation where $\psi^{abcde}$ are real matrices and the Hamiltonian is a pure imaginary matrix, the action of $T$ coincides with complex conjugation $T=K$. Let us consider an eigenstate $\ket{E} = c_i \ket{e_i}$, where $\ket{e_i}$ is a basis that we build with the use of the vacuum and the creation operators $\bar{\psi}^{abcde}$. Then
\begin{gather}
K H_{6,ij} K^{-1} = H_{6,ij}^* = - H_{6,ij},\quad H_{6,ij} c_j = E c_i \Rightarrow  H^*_{6,ij} c^*_j = E c^*_i,\quad H_{6,ij} c^*_j = - E c_i^*.
\end{gather} 
From this one can notice that if the $c_i$ are real then it corresponds to the zero state. Indeed,
\begin{gather}
\braket{E|H|E} = c_i H_{6,ij} c_j = -c_i H_{6,ij}^* c_j = 0.
\end{gather}

To get a symmetry of the Hamiltonian out of the time reversal symmetry, we can combine it with the permutation operator $P_{45}$ to get $T_{45} = T P_{45}$. This operator interchanges two representations of the $A_5$ group. The existence of such a symmetry explains the 6-fold degeneracy of ground state in the numerical studies of the $N=2$ model. The symmetries $A_5$ together with $T_{45}$ form the $S_5$ symmetry group.

With the discrete symmetries of our $q = 6$, $O(N)^5$ symmetric tensor model described above, we are now in a position to find the corresponding random matrix model to describe quantitative properties of the spectrum of the model. This is typically done by mapping our model to a random matrix theory ensemble. There are general rules for choosing the associated ensemble based on the various symmetries of the model \cite{2015Bernevig}. The set of possible ensembles we consider is known as the Andreev-Altland-Zirnbaur (AAZ) ten-fold classification. The symmetries we will use to classify our model are the time reversal symmetry (TRS), and the permutation symmetry described above, $P_{ij}$. As noted above, our Hamiltonian does not posses TRS, like the $q = 6$ SYK model \cite{Maldacena:2016hyu}. In the absence of TRS, we can take $P_{ij}^2 =  + 1$, and we can classify this Hamiltonian as belonging to the AIII ensemble of the AAZ ten-fold classification \cite{Krishnan:2017ztz}. With this classification, we find that the corresponding random matrix ensemble is a chiral Gaussian Unitary Ensemble (chGUE) \cite{Verbaarschot:2005}. We may also use our knowledge of these discrete symmetries to examine the singlet spectrum and its degeneracies, which is done in the following section.

\section{The spectrum of eigenstates of the $O(2)^5$ model}\label{singlets}
In this section, we will study the spectrum of the Hamiltonian (\ref{Hamil}) for $N_i=2$.
The number of different Majorana fermions in this theory is $2^5$, so that there are $2^{16}=65536$ states. 
We can represent each fermion by a gamma matrix of $SO(32)$. We construct the pure real gamma matrices of $SO(32)$ by taking tensor products of Pauli spin matrices, as described in \cite{freedman}. After substituting them into the Hamiltonian (\ref{Hamil}) we obtain a matrix which can be diagonalized using a computer program.

We begin by describing the $SO(N)^5$ invariant states in our theory. They are present only when $N$ is even, and we restrict to this case. In order to count the number of these states, we follow the method of \cite{Klebanov:2018nfp}. We gauge the free theory to get
\beq
\mathcal{S}_G = \int dt \left[\psi_{abcde} \partial_t \psi_{abcde} +  A^1_{a_1 a_2}\psi_{a_1 bcde}\psi_{a_2 bcde}+\ldots\right]. \label{Gauging}
\eeq
The procedure of gauging eliminates all non-singlet states from the spectrum. Indeed, if we calculate the path integral on the circle of the length $\beta$ and first take the integral over the gauge field we get a constraint $J^i_{a b} = 0$ --- the generator of rotations must be equal to zero. After that, the integral over fermions easy to take and we get,
\beq
\int [d\psi]\prod^5_{i=1}[dA^i] e^{i\mathcal{S}} =  \tr_{\rm sing} 1 = N_{\rm singlets}.
\eeq 
If we first calculate the path integral over fermions and gauge the $A^i$ to Cartan subalgebra, where $A^i$ is a skew-symmetric matrix, we get that
\beq
N_{\rm singlets} = 2^{15} \int \prod^5_{i=1} d \Omega^i_{SO(N)} \prod_{k_1=1,\ldots, k_5=1}^N \prod_{\pm} \cos \left[\frac{x^1_{k_1} \pm x^2_{k_2} \pm x^3_{k_3} \pm x^4_{k_4} \pm x^5_{k_5} }{2}\right].\label{thesinglets}
\eeq
Here, $x^i_k, k=1, \ldots, N/2$ and $d\Omega^i_{SO(N)}$ are coordinates and a Haar measure of the $i^{\rm th}$ group. The second product is taken for all possible combinations of the signs. Roughly speaking, the integrand is a character of $SO(N)^5$ and we can decompose it via the characters of the irreducible representations of the group to count the number of the representations. For the case $SO(2)^5$, the integral \eqref{thesinglets} gives 222 singlet states, agreeing with the numerical results. Using the same method, we may count the number of singlet states for models of different ranks. For instance, the $O(2)^4 \times O(4)$ model has 106096 singlets.
\begin{figure}[ht!]
  \centering
\begin{minipage}[c]{0.53\textwidth}
    \includegraphics[width=\textwidth]{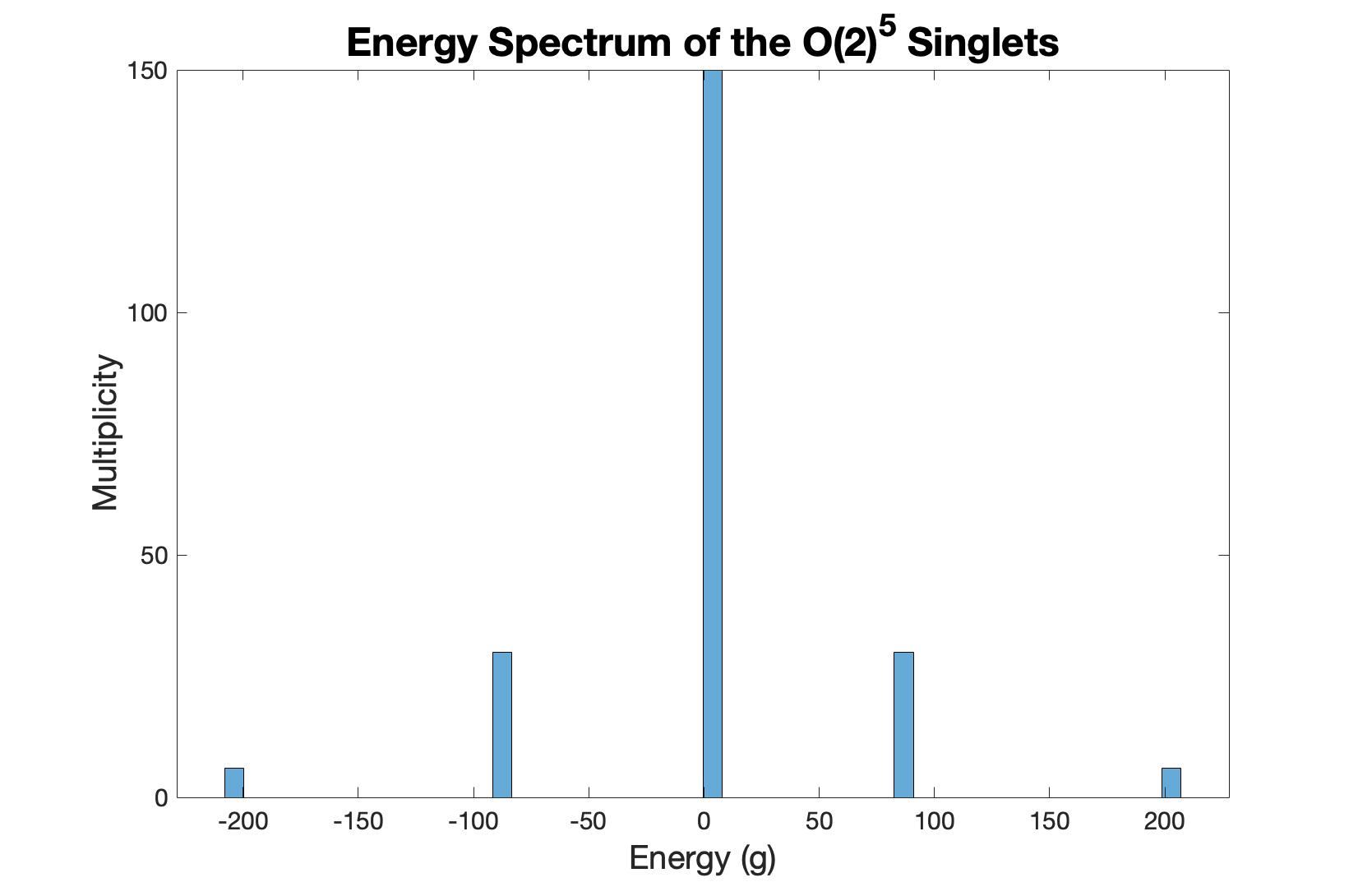}
    \label{fig:singlets}
  \end{minipage} 
  \hfill
 \begin{minipage}[c]{0.45\textwidth}
\begin{tabular}{ |c|c|c| } 
 \hline
Energy&Count&Percent \\ \hline
$-64 \sqrt {10} \approx -202.386$&6&2.70\\ \hline
$-64\sqrt 2\approx  -90.51$&30&13.51\\ \hline
 0&150&67.57\\ \hline
$64 \sqrt 2 \approx 90.51$& 30&13.51\\ \hline
$64 \sqrt {10}\approx 202.386$&6&2.70\\ \hline  
  \end{tabular}
  \end{minipage} 
  \caption{The spectrum of the $SO(2)^5$ invariant states in the $O(2)^5$ tensor model.}
\end{figure}

We can see that the degeneracy of each state of the singlet spectrum for $N = 2$ is a multiple of six. The six-fold degeneracy is explained in section \ref{sub_discrete} by 
the discrete symmetry $S_5$. From the precise numerical eigenvalues we can deduce their exact analytic form: $64\sqrt{10}\approx 202.386$ and
$64\sqrt{2}\approx 90.51$. Eigenvalues expressible in terms of square roots have appeared in other tensor models with low $N$ 
\cite{Krishnan:2018hhu,Klebanov:2018nfp,Krishnan:2018jsp,Pakrouski:2018jcc}.

\begin{table}
	\centering
	\begin{tabular}{ |c|c| } 
		\hline
		Degeneracy&Energy (in units of $g$)
		\\ \hline
		6 & $\pm 64 \sqrt{10}$ \\\hline
		30 & $\pm 64\sqrt{2}$ \\\hline
		32 & $\pm 64\sqrt{42}$ \\\hline
		80 &  $\pm 16\sqrt{18\pm 6\sqrt{5}}$\\ \hline
		80 &  $\pm 16 \sqrt{2 \left(5 \pm \sqrt{21}\right)}$ \\ \hline
		160 & $\pm 32\sqrt{11}$ \\ \hline
		160 & $\pm 16 \sqrt{2 \left(9 \pm \sqrt{57}\right)}$ \\ \hline
		160 & $\pm 16 \sqrt{13 \pm \sqrt{73}}$ \\ \hline
		160 & $ E^6- 8704 E^4+ 15794176 E^2 - 3221225472 =0$ \\ \hline
		160 & $ E^6 - 12800 E^4 + 40960000 E^2 - 805306368=0$\\ \hline
		192 & $E^6-20992 E^4+53215232 E^2 - 1275068416=0$ \\ \hline
		110 & $\pm 128$ \\ \hline
		180 & $\pm 64 \sqrt{3}$\\ \hline
		240 & $\pm 32 \sqrt{10}$ \\ \hline
		320 & $\pm48$ \\ \hline
		320 & $\quad \pm 16 \sqrt{9\pm \sqrt{73}}$ \\ \hline
		480 & $\pm75\sqrt{2}$ \\ \hline
		480 & $\quad \pm \sqrt{519\pm 2 \sqrt{37514}}$ \\ \hline
		808 & $\pm32\sqrt{6}$ \\ \hline
		860 & $\pm 64$ \\ \hline
		992 & $\pm 32\sqrt{3}$\\ \hline
		1120 & $\pm 16\sqrt{2}$ \\ \hline
		1208 & $\pm 32\sqrt{2}$\\ \hline
		1440 & $\pm 16\sqrt{10}$ \\ \hline
		1600 & $\pm 16$\\ \hline
		3200 & $\pm 32$\\ \hline
		31772 & $0$ \\ \hline
	\end{tabular}\caption{The exact spectrum of the $SO(2)^5$ tensor model. The expressions agree with the numerical results up to 11 digits past the decimal.}\label{tab:full_spectrum}
\end{table}

\begin{figure}
  \centering
  \includegraphics[scale=0.25]{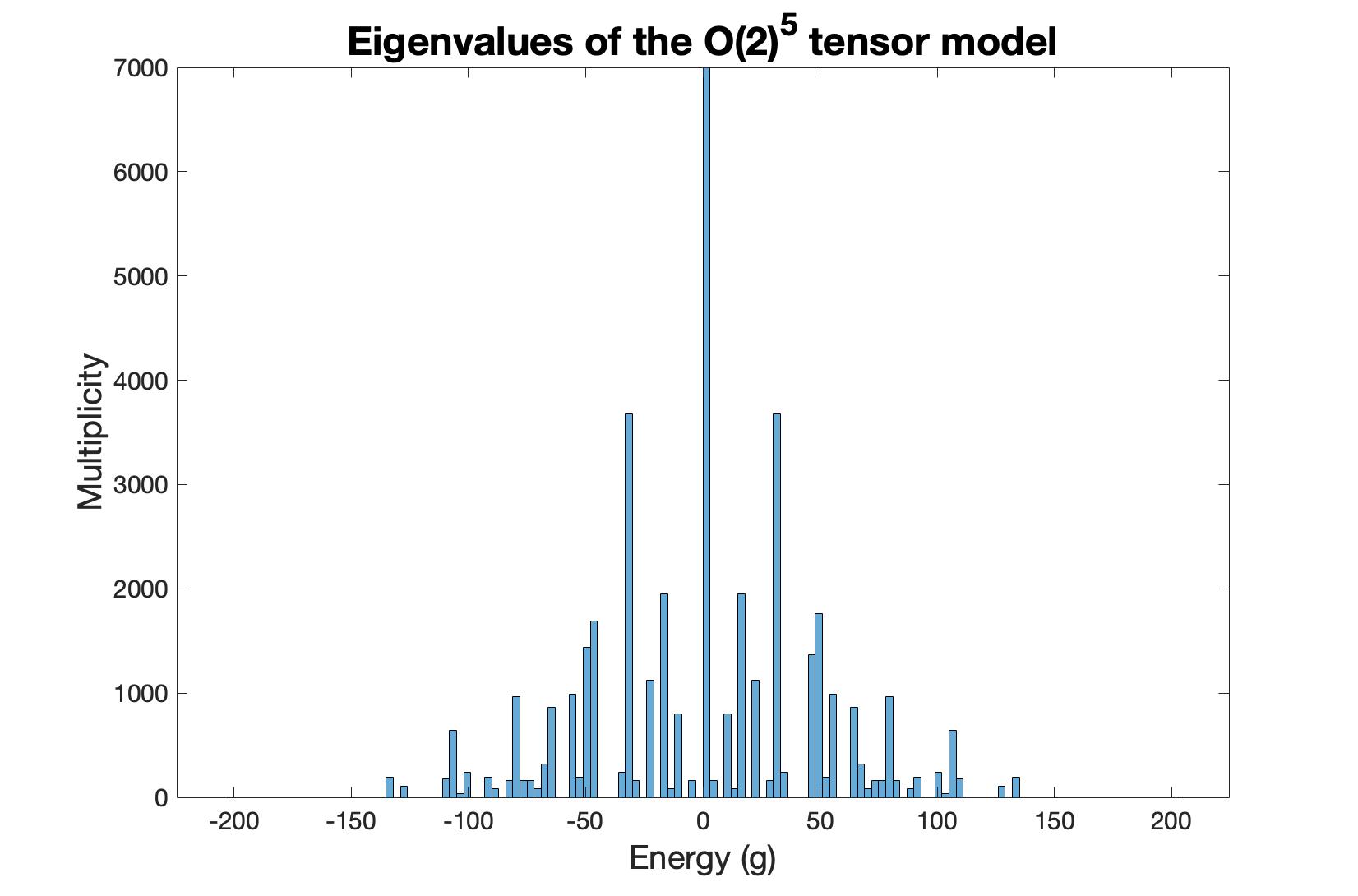}
  \label{fig:q6TM}
  \caption{Energy spectrum of the $O(2)^5$ tensor interaction. There are 31772 zero energy states; not all are displayed.}
\label{fig:tensorspectrum}
\end{figure}

Furthermore, from precise numerical results we have been able to infer the exact expressions for the full spectrum of the $O(2)^5$ tensor model. The energies are found to be roots of even polynomial equations up to order $6$. This is presumably due to the fact that the various symmetries of $H$ allow for mixing of at most six states.
The polynomials have only even powers because they must be invariant under $E\rightarrow -E$,  which follows from the fact that $H\rightarrow -H$ under exchange of any two colors. 
The results are displayed in fig. \ref{tab:full_spectrum}. Most of the eigenvalues may be expressed in terms of square roots or nested square roots, which were seen in other tensor model spectra
\cite{Krishnan:2018hhu,Klebanov:2018nfp,Krishnan:2018jsp,Pakrouski:2018jcc}. 
The remaining $18$ energies are given by the roots of three distinct even sixth-order polynomials.  One of the equations is 
\begin{equation}
E^6- 8704 E^4+ 15794176 E^2- 3221225472 =0\ .
\end{equation}
Its six solutions are given in terms of $\xi=\sqrt[3]{5023+324 i \sqrt{533}}$ as follows: 
\begin{gather}
E_{1,2}= \pm 16 \sqrt{\frac{1}{3} \left(34+\frac{433}{\xi}+\xi\right)}\approx \pm 79.1523 \notag\\
E_{3,4}=\pm\sqrt{\frac{8704}{3}-\frac{55424}{3 \xi}+\frac{55424 i}{\sqrt{3}\xi}-\frac{128}{3}
	\xi-\frac{128 i}{\sqrt{3}}\xi} \approx \pm 46.9662,\notag\\
E_{5,6}=\pm  \sqrt{\frac{8704}{3}-\frac{55424}{3 \xi}-\frac{55424
		i}{\sqrt{3} \xi}-\frac{128}{3} \xi+\frac{128 i }{\sqrt{3}} \xi} \approx \pm 15.2673.
\end{gather}
The roots of the other sixth-order polynomials may be expressed analogously.
The total number of states listed in table \ref{tab:full_spectrum} adds up to $65536= 2^{16}= 2^{N^5/2}$, so it contains the full spectrum, which is shown in fig. 
\ref{fig:tensorspectrum}.

Due to the Poincar\'e recurrence \cite{arnol2013mathematical}, one would expect that any state could return arbitrarily close to the initial state after a sufficient amount of time. One may wonder how to calculate such a time. To do this, we must consider an arbitrarily chosen initial state, which can be decomposed in terms of the eigenstates,
\begin{gather}
\ket{\psi}=\sum_n c_n\ket{E_n},\quad \ket{\psi(t)}= e^{-i H t}\ket{\psi} = \sum_n e^{-i E_n t}c_n\ket{E_n}.
\end{gather}
It follows that the distance between these two states is
\begin{gather}
\left| \ket{\psi(t)} - \ket{\psi}\right|^2 = \sum_n |c_n|^2\left(1-\cos (E_n t)\right). \label{Poirec}
\end{gather}
And, if for any $\eps>0$, there exists a time $t_{\rm rec}$ such that $\left|1-\cos(E_n t)\right| < \eps$, the state $\ket{\psi(t_{\rm rec})}$ is arbitrarily close to the initial state $\ket{\psi}$. The Poincar\'e recurrence theorem guarantees the existence of such a time, but one may wonder how to find it explicitly. Fortunately, if the exact expression for the energies $E_n$ are known, the Lenstra-Lenstra-Lov\'asz (LLL) lattice basis reduction algorithm \cite{ZHANG20173373} may be used to calculate this time. Namely, the condition \eqref{Poirec} for the $t_{\rm rec}$ can be rewritten in the following form. We are looking for the number $q_{\rm rec}$, such that
\begin{gather} \label{Poicon}
\max_n \big| E_n q - \lfloor E_n q\rfloor \big| < \eps.
\end{gather}
The recurrence time in question is $t_{\rm rec}= \frac{q}{2\pi}$. Now, if one constructs the lattice basis in the form
\begin{gather}
\vec{e}_1 = \left(1, Q E_1, Q E_2,\ldots, Q E_n\right),\notag\\
\left(\vec{e}_i\right)_j = \delta_{i,j},
\end{gather}
and applies the LLL algorithm, the first basis vector will have the form,
\begin{gather}
\vec{b}_1=\left(q, Q\left(q E_1- p_1\right), Q\left(q E_2-p_2\right),\ldots\right),\quad \left(q E_i- p_i\right)< Q^{-\frac{1}{n+1}},
\end{gather}
where $p_i$ are integer numbers. Therefore, the number $q$ found by the LLL algorithm is the required $q$ for the condition \eqref{Poicon}.

Applying this algorithm for the spectrum of our model, we find that the recurrence time is
\begin{gather}
t_{\rm rec}=218516231876133437533409856498158380135794428.3096919112 g^{-1} \approx 2.18*10^{45} g^{-1}, \notag\\
\left|1-\left|\frac{Z(t_{\rm rec})}{Z(0)}\right|\right| < 0.5*10^{-2}.
\end{gather}

\section{Comparison with the $q=6$ SYK model}\label{finite_N}

In this section we calculate the energy spectrum and the spectral form factor of the $N_{\rm SYK} = 26$ , $q = 6$ SYK model
and compare with corresponding results of the $O(2)^5$ tensor model. The $q = 6$ SYK model Hamiltonian is
\begin{equation}
H_{\rm SYK} = i \sum_{1\leq i_1< ...< i_6<N_{\rm SYK} }j_{ i_1... i_6}\psi_{i_1}\psi_{i_2}...\psi_{i_6}, \qquad \braket{j_{i_1\ldots i_6} j_{j_1 \ldots j_6}} = J^2 \frac{\delta_{i_1 j_1}\ldots \delta_{i_6 j_6}}{N_{\rm SYK}^5}\ .
\end{equation}
In this case there are $2^{13}=8192$ states, and each fermion is assigned to a gamma matrix of $SO(26)$.

\begin{figure}
  \centering
   \includegraphics[scale=0.25]{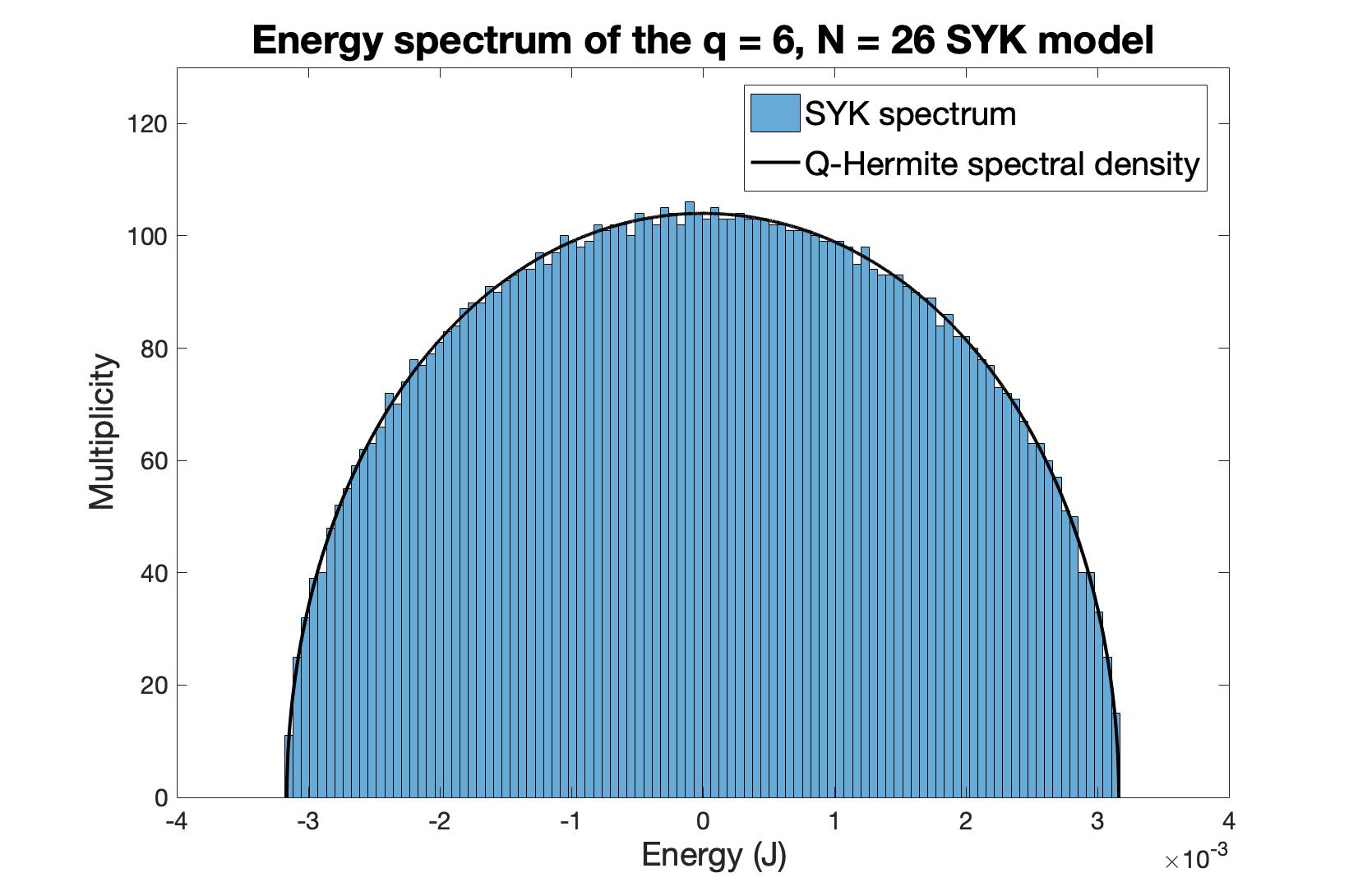} 
   \label{fig:q6SYK}
  \caption{The energy spectrum of the $q=6$ SYK model with $N_{\rm SYK} = 26$, averaged over 49 samples.}
\label{fig:q6SYK}
\end{figure}

In fig. \ref{fig:tensorspectrum}, we can see that there are large energy gaps in the tensor model, whereas the SYK model has a much denser spectrum and displays a near semi-circular distribution of eigenvalues that is characteristic of random matrices. Upon examining the energy spectrum, we can see the $E \rightarrow -E$ symmetry in the $q=6$ model due to the time-reversal symmetry, which is not present in the $q = 4$ SYK model. We provide a fit for the energy spectrum as shown in fig. \ref{fig:q6SYK}. This fit is the spectral density that corresponds to the $Q$-Hermite polynomial with $Q$ equal to a combinatorial factor, $\eta$, that encodes the suppression of crossing diagrams in the Wick contractions of gamma matrices. The suppression factor is derived in \cite{Garcia-:2017},
\begin{equation}
\eta   = {N_{\rm SYK}  \choose q}^{-1}\sum_{p = 0}^q(-1)^p{q \choose p}{N_{\rm SYK} -q \choose q-p}. \\
\end{equation}
The $Q$-Hermite spectral density, $\rho_{QH}(E)$, is the following \cite{Cotler:2016fpe, Garcia-:2017, BAGRETS},
\begin{equation}
\rho_{QH}(E)  = A\sqrt{1-\left(E/E_0\right)^2}\prod_{k=0}^\infty \left[ 1-\left(2\frac{E}{E_0}\right)^2\frac{1}{1+\eta^k+\eta^{-k}} \right]\label{QHerm}
\end{equation}
where $A \approx 104$ is the normalization constant, which imposes that the total number of states is equal to $2^{N_{\rm SYK} /2} = 8192$, $E_0 \approx -0.0032$ $J$ is the ground state energy, and $\eta \approx -0.0072$ is the suppression factor. The spectral density, \eqref{QHerm}, is calculated in the double scaled limit, where $N_{\rm SYK}  \rightarrow \infty$, $q \rightarrow \infty$, and $q^2/N_{\rm SYK} $ fixed. We can see that there is strong agreement with the Q-hermite polynomial and the $q = 6$ SYK energy spectrum, which indicates that this model is a very good approximation of the double scaled limit.

\begin{figure}
  \centering
  \begin{subfigure}{\textwidth}
    \includegraphics[width=\textwidth]{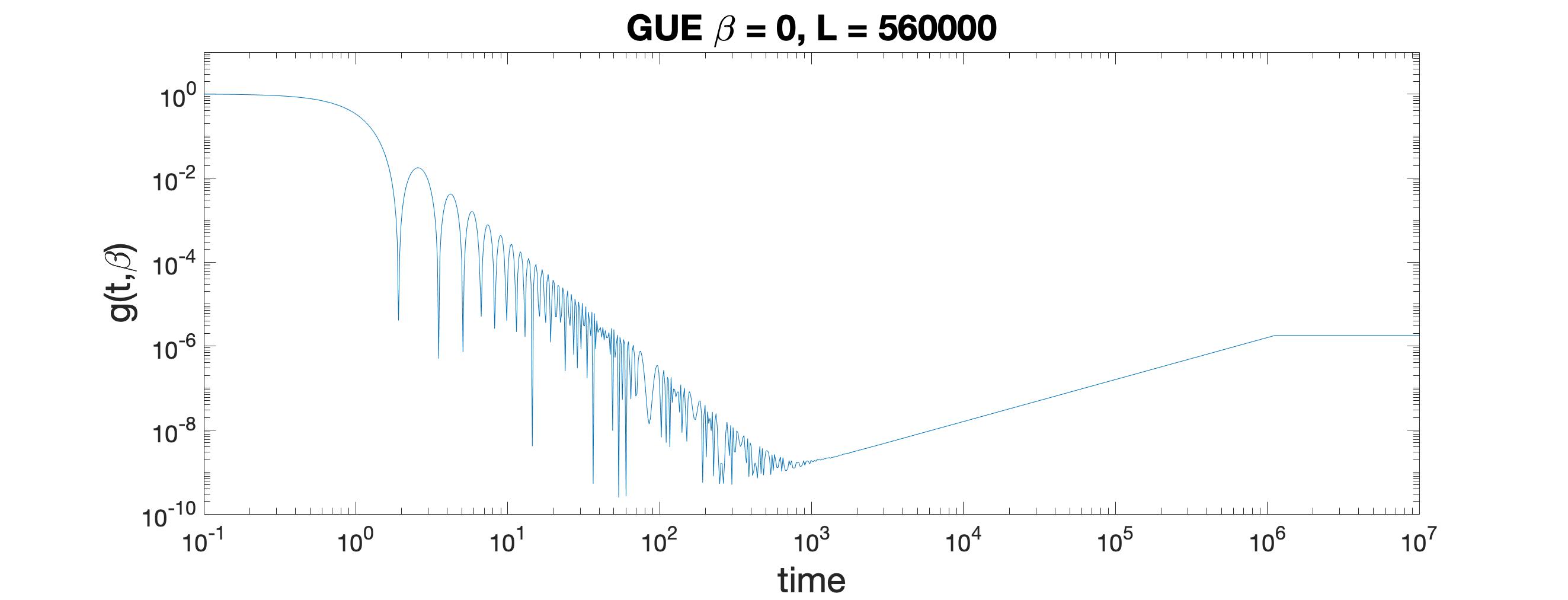}
  \end{subfigure} 
    \hfill
  \begin{subfigure}{\textwidth}
    \includegraphics[width=\textwidth]{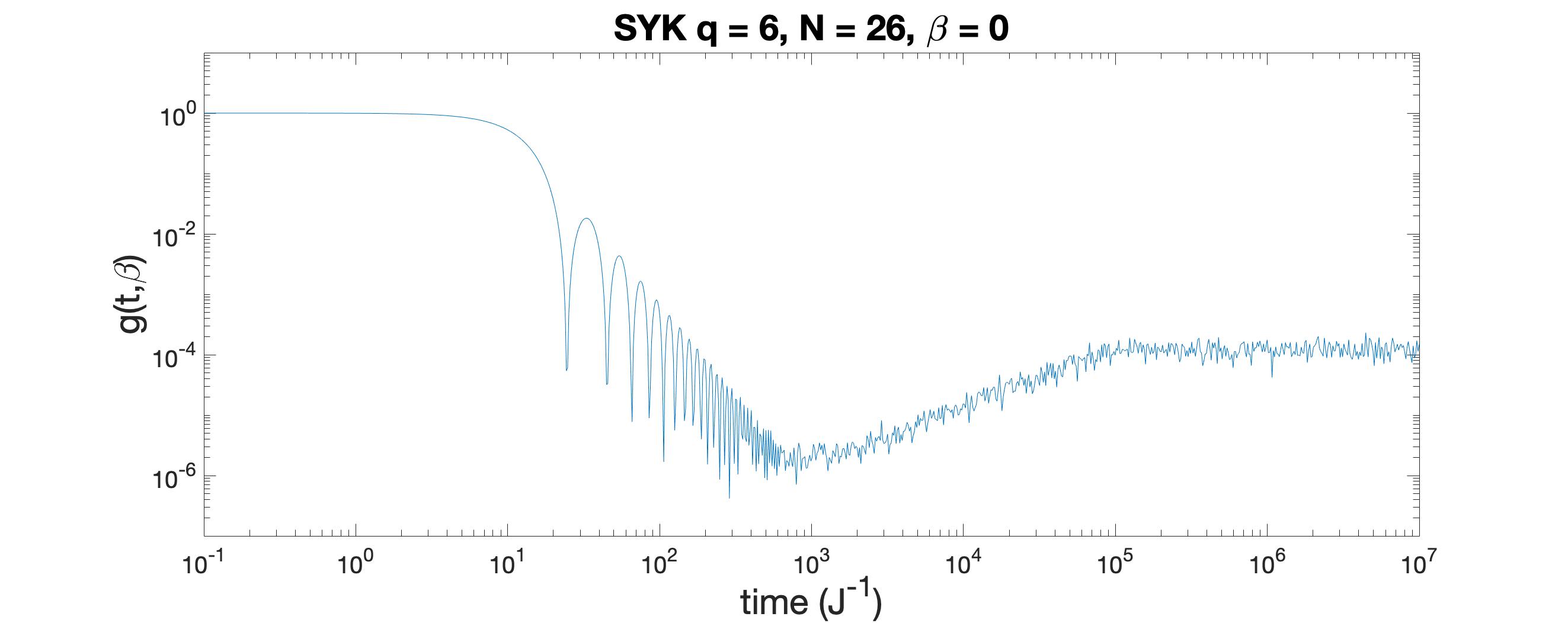}
  \end{subfigure}
  \hfill
  \begin{subfigure}{\textwidth}
    \includegraphics[width=\textwidth]{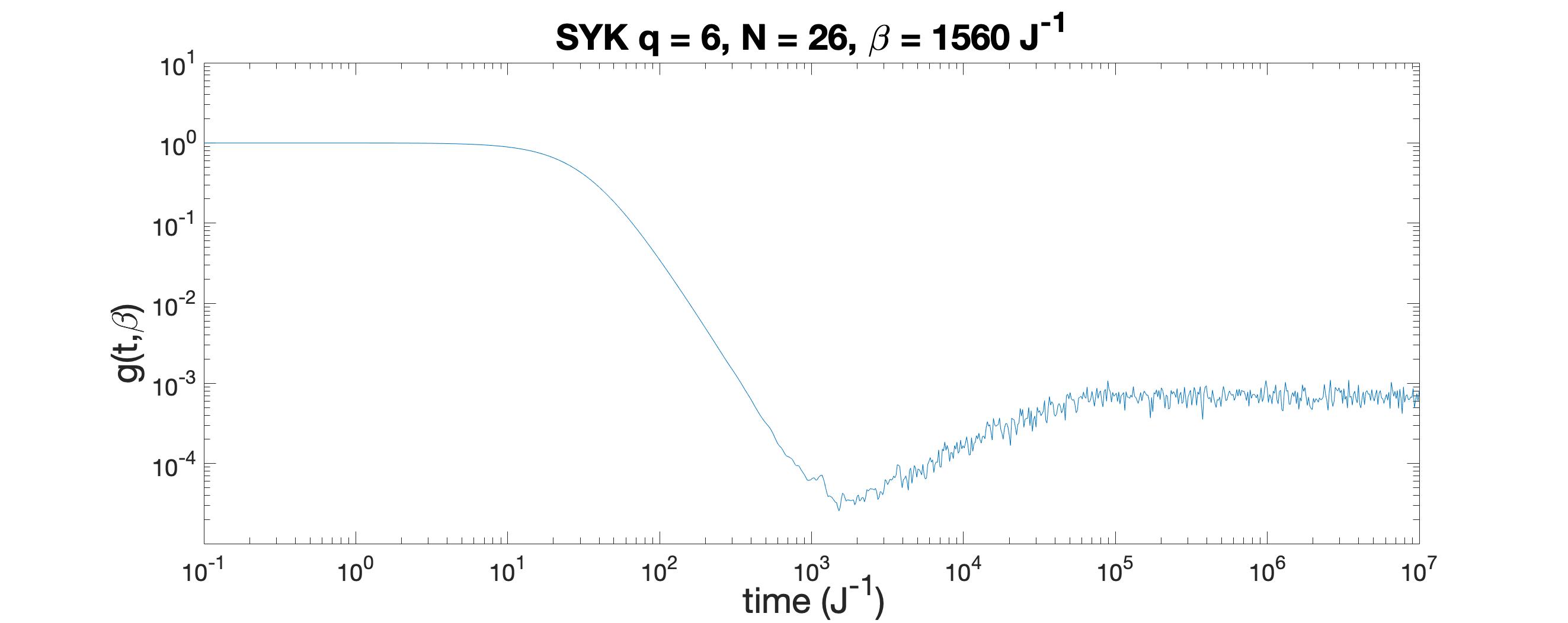}
  \end{subfigure}
   \hfill
  \caption{Top: SFF for the gaussian unitary ensemble (GUE) associated with the $q = 6$, $N_{\rm SYK}  = 26$ SYK model at $\beta = 0$. Middle: SFF for the $q = 6$, $N_{\rm SYK}  = 26$, $\beta = 0$ SYK model averaged over 49 samples. Bottom: SFF for the $q = 6$, $N_{\rm SYK}  = 26$, $\beta = 1560$ $J^{-1}$ SYK model averaged over 49 samples.}\label{fig:SYKdrp2}
\end{figure}

\begin{figure}
  \centering
  \begin{minipage}[b]{\textwidth}
    \includegraphics[width=\textwidth]{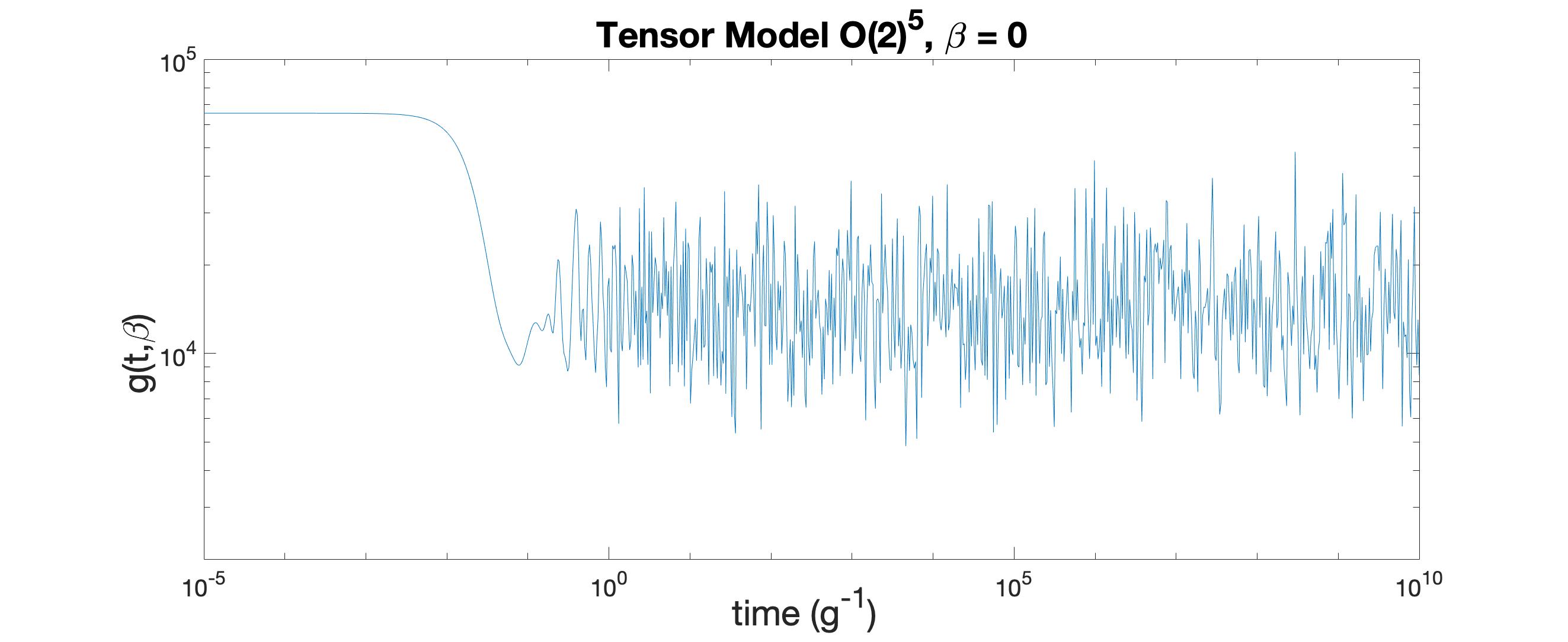}
  \end{minipage}
    \hfill
  \begin{minipage}[b]{\textwidth}
    \includegraphics[width=\textwidth]{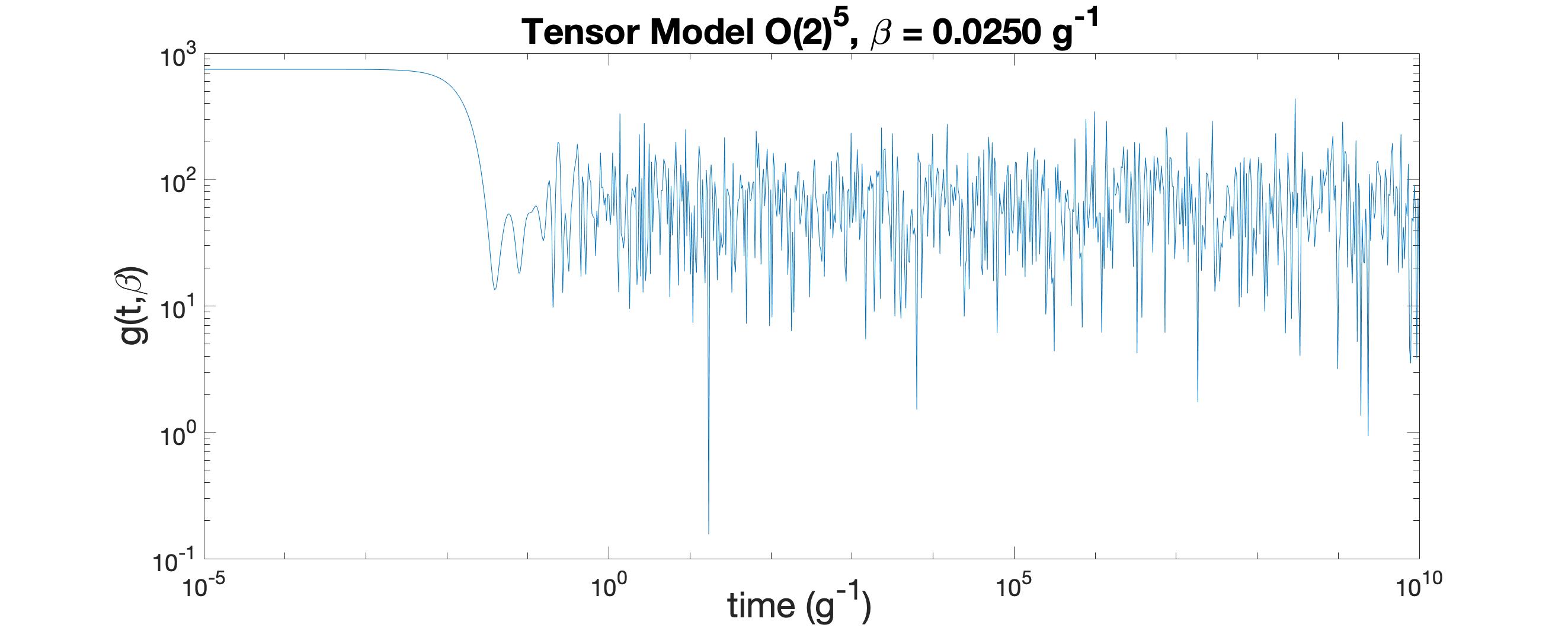}
  \end{minipage}
  \hfill
  \begin{minipage}[b]{\textwidth}
    \includegraphics[width=\textwidth]{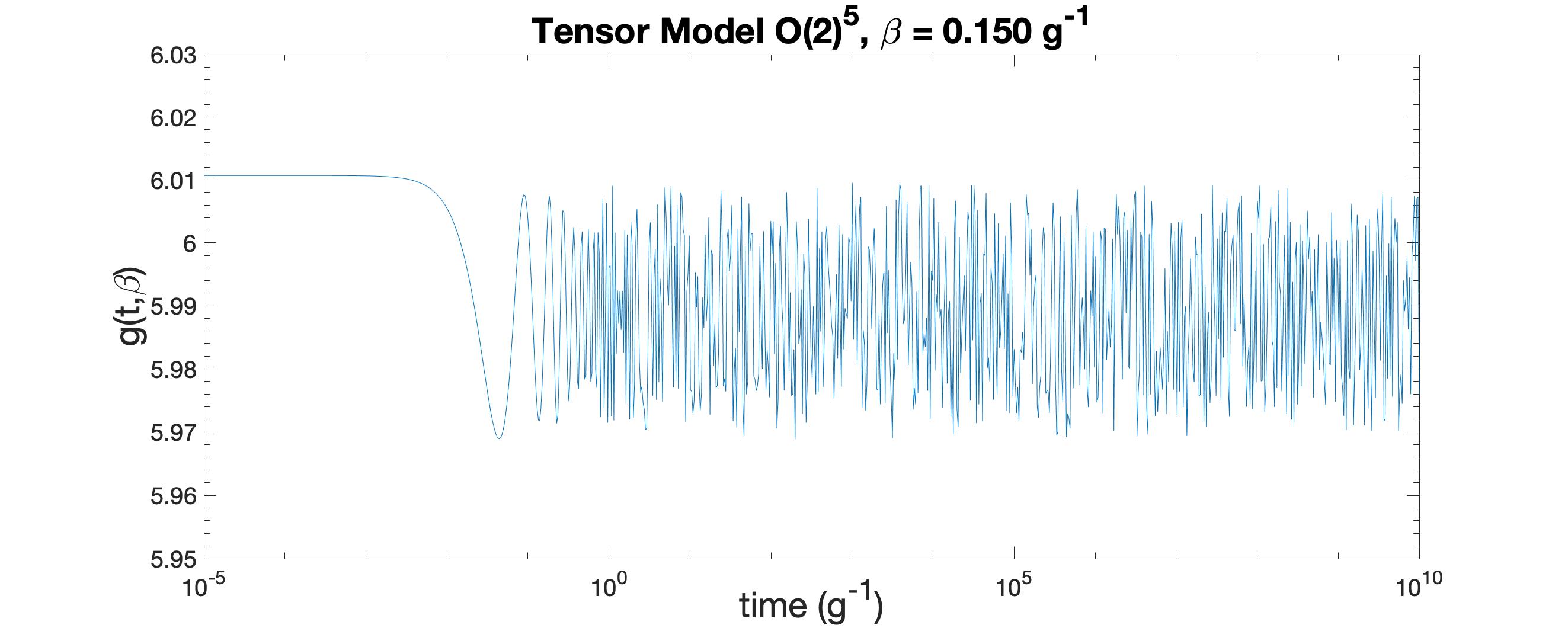}
  \end{minipage}
   \hfill
  \caption{SFF for the $O(2)^5$ tensor model for three values of $\beta$. Top: $\beta = 0$, middle: $\beta = 0.0250$ $g^{-1}$, and bottom: $\beta = 0.150$ $g^{-1}$. Note that the fluctuations for the bottom subfigure are much smaller than the two above --- this is because the SFF is calculated at a lower temperature.}\label{fig:TM1drp20}
\end{figure}

Additionally, we can examine and compare the spectral form factor (SFF) for the SYK and tensor models (similar calculations in tensor models with $q=4$ were performed in 
\cite{Krishnan:2016bvg,Krishnan:2017txw}). The SFF is a measure of the discreteness of the energy spectrum and can be defined as
\cite{Cotler:2016fpe,Garcia-Garcia:2016mno}
\begin{equation}
\begin{split}
&g(t,\beta) = |Z(t,\beta)|^2/Z(\beta)^2,\quad \text{where}\quad
Z(t,\beta) = \text{Tr}(e^{-\beta H-iHt})\ .
\end{split}
\end{equation}
In fig. \ref{fig:SYKdrp2} and fig. \ref{fig:TM1drp20}, we display plots of the SFF for the $q=6$ SYK and tensor models. For comparison, we have also plotted the SFF of the corresponding random matrix theory (RMT) ensemble, which is determined by the value of $N_{\rm SYK} $ mod 8 \cite{Cotler:2016fpe}. In our case, we plot for $N_{\rm SYK}  = 26$, which is associated with the gaussian unitary ensemble (GUE). The SFF for the GUE that we have plotted is calculated in \cite{Cotler2017}, and we have included the result at infinite temperature below:
\begin{equation}
g(t)_{GUE} = L^2\left(\frac{J_1(2t)}{t}\right)^2+L-L\times
     \begin{cases}
       1-\frac{t}{2L}, &t < 2L\\
       0, & t >2L \\
     \end{cases}
\end{equation}
$J_1(t)$ is the Bessel function of the first kind, and contributes to the early time oscillations of the GUE. $L$ sets the size of the ensemble of random hermitian matrices, and is related to the plateau time as $t_p = 2L$.

We can see that the SFF for the SYK model has the same features of the corresponding RMT ensemble, indicating properties of quantum chaos; in particular, the dip-ramp-plateau structure is present (see fig. \ref{fig:SYKdrp2}). Some of these properties are more difficult to see in the tensor model because the gaps in the energy spectrum are sizable for the available value of $N$. However, we can notice a dip and plateau structure in our tensor model, which suggest signs of chaotic behavior, but there is no obvious ramp (see fig. \ref{fig:TM1drp20}).

Despite clear differences in the finite $N$ behavior of the tensor model and SYK model, we find that the large $N$ solutions of the two models are identical. Before solving the large $N$ models, we will discuss higher $q$ tensor models followed by the large $N$ limit and the melonic dominance of our tensor model.

\section{Tensor models with $q>6$ }\label{higher_q}

We begin with a discussion of $q = 8$, where the Majorana fermion tensor is of rank seven, and the model has $O(N)^7$ symmetry. In a ``generalized tetrahedral" interaction vertex, every two tensors have exactly one index in common. 
In contrast to the $q = 6$ case, there are six distinct such $q = 8$ interactions \cite{Ferrari:2017jgw,Gubser:2018yec}. However, only one of these interactions has the property that it stays connected whenever any $5$ colors are erased. 
This is the maximally single-trace (MST) vertex in the terminology of \cite{Ferrari:2017jgw}, and we will show that in the Majorana model it produces a Hamiltonian which is fully antisymmetric under interchange of the $O(N)$ groups. 
The problem of finding the MST interactions is equivalent to the problem of finding the perfect 1-factorization of the complete graphs \cite{midori88}. There are two classes where the existence of the perfect 1-factorizations has been proven: for graphs with $p+1$ vertices or $2p$ vertices, where $p$ is an odd prime number. 

\begin{figure}
\centering
{\includegraphics[width=0.3\textwidth]{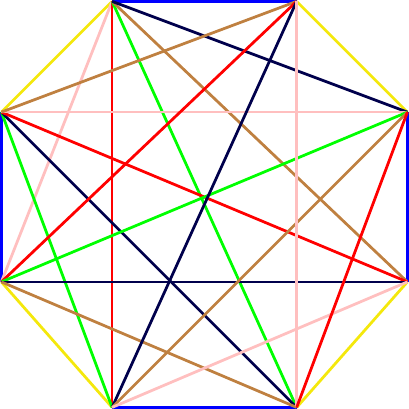}}
\caption{\label{fig:q8} A graphical representation of the unique maximally single-trace tensor interaction for $q = 8$. It stays connected when any $5$ out of the $7$ colors are erased.}
\end{figure}

The $q = 8$ MST interaction is shown in fig. \ref{fig:q8}. This interaction is called the canonical coloring \cite{Gubser:2018yec}; this means that if we erase any set of $5$ colors, we are left with an octagon composed of alternating colors. We can show the antisymmetry of this fermionic interaction as follows. 
Let us erase all colors except for groups $O(N)_a$ and $O(N)_b$ to get,
\begin{gather}
H_8 = 
\psi^{a_1 b_1 c_1 d_1  e_1f_1g_1}\psi^{a_1 b_2 c_2 d_2 e_2f_2g_2} \psi^{a_2 b_1 c_3 d_3 e_3f_3g_2} 
\psi^{a_2 b_3 c_1 d_4 e_2f_4g_3} \\ \psi^{a_3 b_4 c_3 d_1 e_4f_2g_3} \psi^{a_3 b_2 c_4 d_4 e_1f_3g_4} \psi^{a_4 b_3 c_2 d_3 e_4f_1g_4} \psi^{a_4 b_4 c_4 d_2 e_3f_4g_1} \longrightarrow\notag\\
\longrightarrow
H_2= 
\psi^{a_1 b_1}\psi^{a_1 b_2} \psi^{a_2b_1} 
\psi^{a_2 b_3} \psi^{a_3 b_4} \psi^{a_3 b_2 }
 \psi^{a_4 b_3} \psi^{a_4 b_4}.
\end{gather}
Now let us exchange the $O(N)_a$ and $O(N)_b$ groups of $H_2$ to get,
\begin{equation}
\begin{split}
H_2'&= 
\psi^{a_1 b_1}\psi^{a_2 b_1} \psi^{a_1b_2} 
\psi^{a_3 b_2} \psi^{a_4 b_3} \psi^{a_2 b_3 }
 \psi^{a_3 b_4} \psi^{a_4 b_4}\\
 &= - \psi^{a_1 b_1}\psi^{a_1 b_2} \psi^{a_2b_1} 
\psi^{a_2 b_3} \psi^{a_3 b_4} \psi^{a_3 b_2 }
 \psi^{a_4 b_3} \psi^{a_4 b_4} = -H_2.
 \end{split}
\end{equation}

\begin{figure}
	\centering
	\begin{minipage}{.5\textwidth}
		\centering
		{\includegraphics[width=0.45\textwidth]{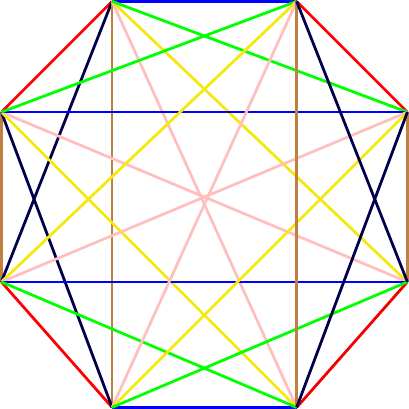}}
\end{minipage}%
	\begin{minipage}{.5\textwidth}
	\centering
	{\includegraphics[width=0.45\textwidth]{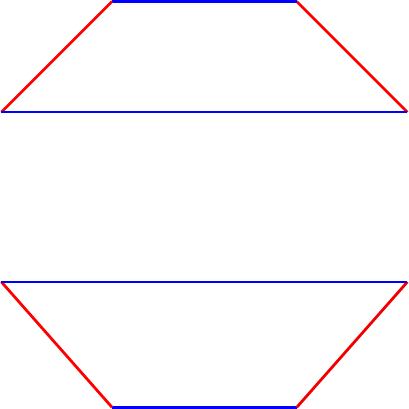}}
	\end{minipage}
\caption{\label{fig:q8_notantisym} The graphical representation of a $q = 8$ tensor interaction which is not maximally single-trace. If we erase all but the blue and red stands, the graph becomes disconnected.}
\end{figure}

This is in contrast to the other $q = 8$ interactions that satisfy the constraint that one index is shared among any two pairs of fermions, all of which are provided in fig. 2 of \cite{Gubser:2018yec}. We give an example of a non-MST interaction in fig. \ref{fig:q8_notantisym}, corresponding to fig. 2,a in \cite{Gubser:2018yec}. When we erase all but two colors, we are left with two disconnected diagrams, which means this interaction is symmetric under exchange of these two colors.

Let us now comment on the $q=8$ MST interaction. Since there is no $i$ in this interaction, we have the time reversal symmetry. The $E \rightarrow -E$ symmetry comes from the antisymmetry under the exchange of two gauge groups. This interaction is melonic and scales as $g^2N^{\frac{(q-1)(q-2)}{2}}=g^2N^{21}$, following the arguments in section \ref{melonic_dominance}. In section \ref{large_N}, we will calculate the scaling dimensions of the bilinears of this model, and also include the result for general $q$ tensor models.

We will define the group of coloring automorphisms, which will be used in calculating the propagator. One can think of a coloring automorphism as a permutation of the vertices of the interaction graph in a way that preserves the colors of the edges. Paper \cite{Gubser:2018yec} explores these symmetries in more detail, and shows that the group of coloring automorphisms is $\mathbb{Z}_2^n$. Furthermore, \cite{Gubser:2018yec} proves that for $q = u2^v$, $u$ odd, melonic tensor models, the group of coloring automorphisms, which we will denote as $Aut$, can be at most $\mathbb{Z}_2^v$ for $u = 1$ and $\mathbb{Z}_2^{v-1}$ for $u >1$.

There are six distinct $q=8$ interactions that satisfy the constraint that each pair of Majorana fermions has a single index contraction. The difference between them is the order of the coloring automorphism group, which is taken into account in \eqref{DSequation}. The more symmetry our interaction has, the larger the order of the automorphism group will be. It follows that the $q = 8$ fully symmetric diagram has the largest group order, with $Aut = \mathbb{Z}_2^3$ \cite{Gubser:2018yec}.  As noted in section \ref{large_N}, the $|Aut|$ factor cancels out in the spectra calculation. 
\begin{figure}
\centering
{\includegraphics[width=0.45\textwidth]{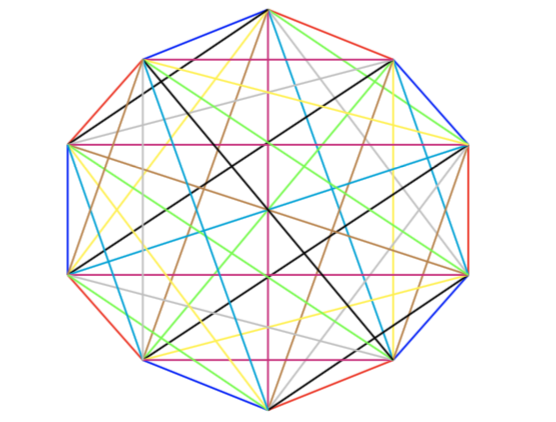}}
\caption{\label{fig:q10} The graphical representation of the maximally single-trace tensor interaction for $q = 10$.}
\end{figure}

The number of possible ``generalized tetrahedral" interactions increases very rapidly with $q$ \cite{Ferrari:2017jgw,Gubser:2018yec}: for $q=8$ it is $6$, for $q = 10$ it is 396, and for $q = 12$ it is $526,915,620$. However, at least for $q=8$ and $10$ the maximally single-trace vertex, or perfect one-factorization, is unique \cite{Ferrari:2017jgw}.\footnote{The smallest value of $q$ where the MST vertex is not unique is $12$. We thank Fidel Schaposnik Massolo for informing us of this and providing a reference, \cite{colbourn2010crc}.}
For $q = 10$, the MST vertex is shown in fig. \ref{fig:q10} (see also fig. 5 of \cite{Ferrari:2017jgw}). 

\section{Melonic dominance for maximally single-trace interactions}\label{melonic_dominance}

\begin{figure}
	\centering
	\begin{minipage}{.5\textwidth}
		\centering
		{\includegraphics[width=0.25\textwidth]{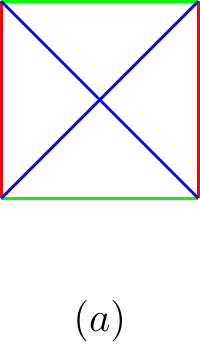}}
	\end{minipage}%
	\begin{minipage}{.5\textwidth}
			\centering
			{\includegraphics[width=0.35\textwidth]{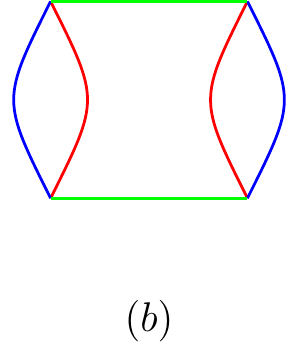}}
	\end{minipage}
		\caption{Diagrams of the tetrahedral operator (a) and one of the pillow operators (b).}
	\label{H_tetra_pillow}
\end{figure}

In this section we discuss the structure of Feynman diagrams contributing at leading order in $N$; they are often called the maximal diagrams. First, let us recall the tensor model with $O(N)^3$ symmetry, corresponding to $q=4$. This model has single-sum interaction vertices of either the tetrahedron type or the pillow type 
\cite{Carrozza:2015adg, Klebanov:2016xxf}. A representative of the latter is  
\begin{align}
H_p = \frac{g_p} {4} \psi^{a_1 b_1 c_1 }\psi^{a_1 b_1 c_2 } \psi^{a_2 b_2 c_1 } 
\psi^{a_2 b_2 c_2}\ , \label{pillowgreen}
\end{align}
which is illustrated in fig. \ref{H_tetra_pillow},b.

Let us study the vacuum Feynman graphs of this theory and take turns erasing the strands of a given color. For the maximal graphs, the remaining double-line diagrams are planar, since increasing their genus decreases the number of loops \cite{Carrozza:2015adg, Klebanov:2016xxf}.
If such a double-line diagram has $n$ separate connected components, then the Euler theorem states that the number of index loops is given by
\begin{equation}
f_{rb} = 2n_{rb} + v_t,\quad \text{and}\quad f_{rg,bg}=2 n_{rg,bg}+v_t+v_p\ ,
\end{equation}
where $v_t$ and $v_p$ are the numbers of the tetrahedral and pillow vertices, respectively. Since the pillow vertex (\ref{pillowgreen}) becomes disconnected when the green strands are erased, we find that the number of separate components of the red-blue graph satisfies
\begin{equation}
n_{rb}\leq 1+ v_p\ .
\end{equation} 
On the other hand, the tetrahedral vertex stays connected when red or blue strands are erased, so that $n_{rg}=n_{bg}=1$. These numbers are independent of $v_t$ because the tetrahedral vertex stays connected when any color is erased
\begin{align}
& f_{rb}= f_r + f_b\leq 2 + v_t + 2 v_p 
\ , \nonumber \\
& f_{rg}= f_r + f_g= 2 + v_t + v_p \ , \nonumber \\
& f_{bg}= f_b + f_g= 2 + v_t + v_p \ .
 \label{strands}
\end{align}
Adding these equations, we find that the {\it maximum} total number of closed loops is
\begin{equation}
f_r+ f_b+ f_g= 3 + \frac{3}{2} v_t + 2 v_p
\ .
\end{equation} 
This means that the maximum weight of a graph is $N^3 \lambda_t^{v_t} \lambda_p^{v_p}$. Here
\begin{equation}
\label{tetrapillowscaling}
\lambda_t = g_t N^{3/2}\ , \qquad \lambda_p = g_p N^{2}
\end{equation} 
are the quantities which must be held fixed to achieve a smooth large $N$ limit. These scalings apply to any rank-$3$ tensor theory with $O(N)^3$ symmetry and 
quartic interactions \cite{Carrozza:2015adg,Klebanov:2016xxf,Giombi:2017dtl}.\footnote{In the special case of quantum mechanics of Majorana fermions $\psi^{abc}$, 
the pillow operators are simply the quadratic Casimir invariants of the $O(N)$ groups. It is possible to show that their maximal  
values in the Hilbert space are of order $N^5$. This means that the energy shift for such states due to the pillow operator is $\sim g_p N^5 \sim \lambda_p N^3$. The fact that this scales as the number of degrees of freedom, $N^3$, is a confirmation that the scaling (\ref{tetrapillowscaling}) is correct.} 

The discussion above shows that the simplest melonic large $N$ limit applies to the $g_p=0$ model which has a purely tetrahedral interaction. The tetrahedron vertex stays connected when
the strands of one color are erased and becomes a connected double-line vertex, which 
is found in the $O(N)\times O(N)$ symmetric matrix model with a single-trace interaction $g_t \tr (M M^T)^2$. In the $O(N)^3$ model, the tetrahedral vertex is the unique quartic vertex which is maximally single-trace.
Let us now perform a similar analysis in the large $N$ limit of $O(N)^{q-1}$ symmetric tensor models corresponding to higher even values of $q$. To achieve the simplest large $N$ limit we will consider only the maximally single-trace interaction vertices \cite{Ferrari:2017jgw}, which stay connected whenever any $q-3$ colors are erased. The unique such interaction vertex for $q=6$, (\ref{classH}), is shown in fig. \ref{fig:tikzsix}, for $q=8$ in fig. \ref{fig:q8}, and for $q=10$ in fig. \ref{fig:q10}. When colors $i$ and $j$ are left, the double-line vertex is of the kind found in a   $O(N)\times O(N)$ symmetric matrix model with the single-trace interaction $ g \tr (M M^T)^{q/2}$. Since this interaction is single-trace, the two-color graph may be drawn on a {\it connected} Riemann surface of genus $g_{ij}$, and we have the  
constraint
\begin{equation}
f_{ij} + v - e =  2 - 2 g_{ij} 
\ ,
\end{equation}
where $e$ and $v$ are the total numbers of the edges and the vertices.  
Since the graphs may be non-orientable, the possible values of the genera, $g_{ij}$, are $0,1/2, 1, \ldots$. Using $e= q v/2$ and summing over all choices of remaining two colors we find
\begin{equation}
\sum_{i < j} f_{ij} = (q-1) (q-2) + {(q-1) (q-2)^2 \over 4} v - 2  \sum_{i < j} g_{ij}
\ .
\end{equation} 
Since 
\begin{equation}
\sum_{i < j} f_{ij} = (q-2)\sum_i f_i = (q-2) f_{\rm total} 
\ ,
\end{equation}
we find
\begin{equation}
f_{\rm total}  = q-1 +{(q-1) (q-2) \over 4} v - \frac{2}{q-2}  \sum_{i < j} g_{ij}\ . \label{HigherQFace}
\end{equation}
The maximum possible weight of a vacuum graph with $v$ vertices, corresponding to all $g_{ij}=0$, is
\begin{equation}
N^{q-1} \lambda^{v}\ , 
\end{equation}
and the large-$N$ limit needs to be taken with
\begin{equation}
\lambda = g N^{(q-1)(q-2)/4}
\end{equation}
held fixed.\footnote{This large-$N$ scaling is the same as in the Gurau-Witten model \cite{Gurau:2009tw,Witten:2016} for $q$ flavors of rank $q-1$ tensors.} We see that the large-$N$ partition function of the $O(N)^{q-1}$ tensor model has the structure
\begin{equation}
\lim_{N\rightarrow \infty} N^{1-q} \ln Z = f(\lambda) \ . 
\end{equation}

\begin{figure}
	\centering
	\begin{minipage}{.4\textwidth}
		\centering
		{\includegraphics[width=0.65\textwidth]{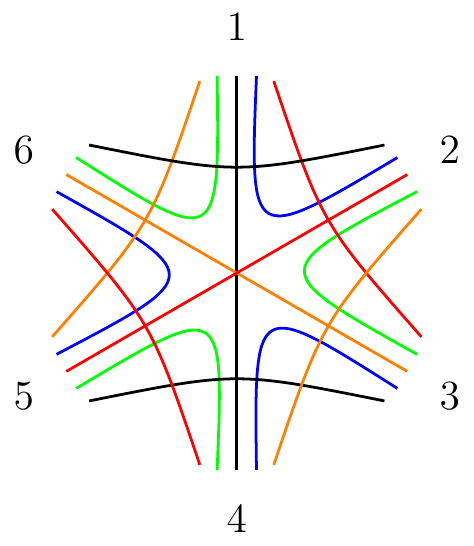}}
	\end{minipage}%
	\begin{minipage}{.4\textwidth}
		\centering
		{\includegraphics[width=0.65\textwidth]{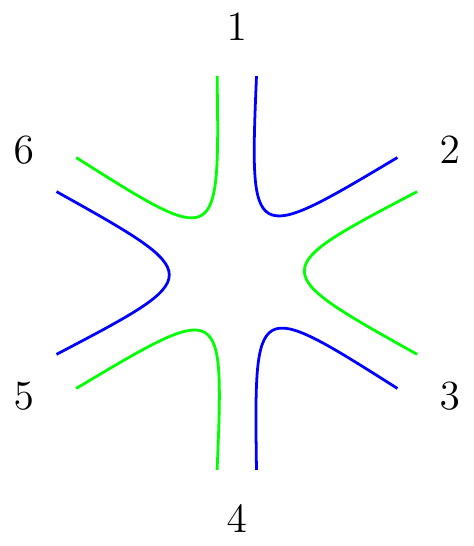}}
	\end{minipage}
	
	\caption{The vertex becomes single-trace if we keep any two colors.}
	\label{forget3color}
\end{figure}

Now we sketch a proof that the model with a maximally single-trace interaction vertex possesses the melonic dominance in the large $N$ limit --- for such an operator,
forgetting any $q-3$ indices leads to a single-trace operator (a diagrammatic representation of this for $q=6$ is shown in fig. \ref{forget3color}). A more rigorous proof, which is however restricted to cases where $q-1$ is prime, was given in \cite{Ferrari:2017jgw}.

\begin{figure}
	\centering
	\includegraphics{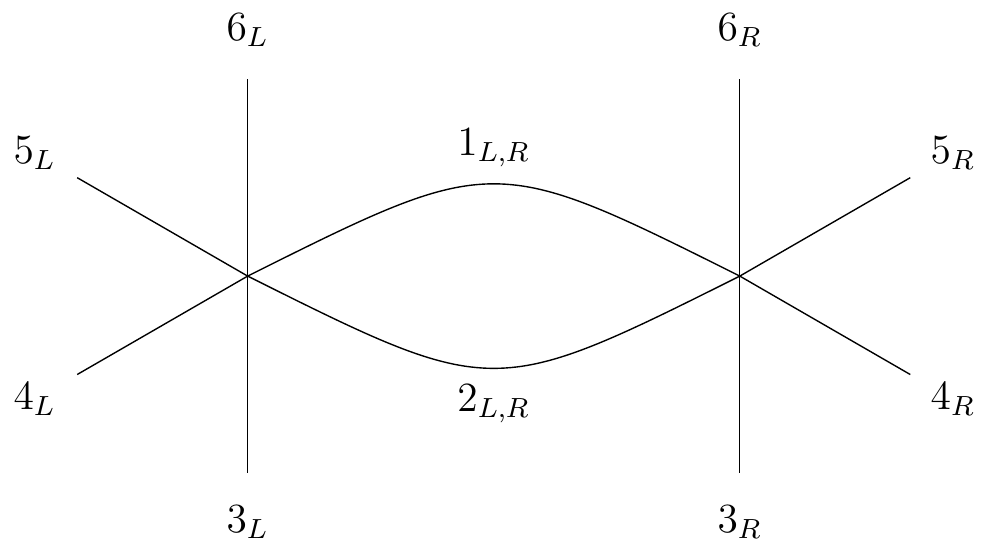}
	\caption{A basis pair of vertices that is connected by a pair of propagators.}
		\label{basispair}
\end{figure}

\begin{figure}
	\centering
 \includegraphics[scale=1]{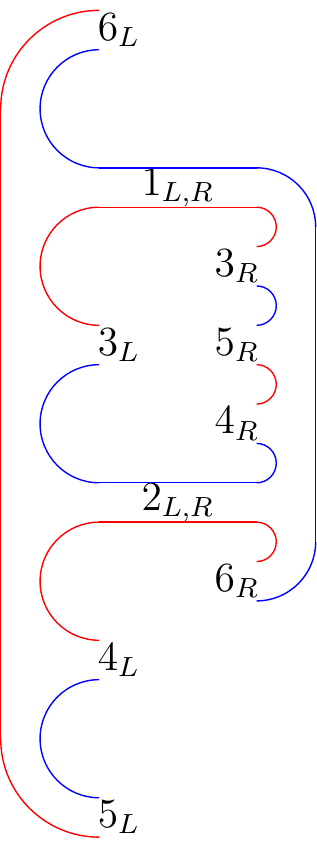}
		\caption{Because we consider a maximally single-trace operator, we can erase all except two colors and have a single-trace vertex. If they are connected to each other by two propagators, then the most general structure could be only the one shown in this figure. For the output $3_L$ in this case we assign the number $r_3=3$.}
	\label{132fig}
\end{figure}

As we have shown, the graphs giving the leading contribution in the large $N$ limit have $g_{ij}=0$, i.e., any choice of the double-line graph is planar. In this case
we find
\begin{gather}
f_{\rm total} = q-1 + \frac{(q-1)(q-2)}{4}v\ .
\end{gather}
Let us show that there is a loop passing through only 2 vertices and use the strategy analogous to that in the $q=4$ case \cite{Klebanov:2016xxf}. Let $f_r$ denote the number of loops passing through $r$ vertices. Since there are $\frac{q(q-1)}{2}$ strands meeting at every vertex, we find the sum rules
\begin{gather}
\sum f_r = f_{\rm total}\ ,\qquad \sum_r r f_r = \frac{q(q-1)}{2} v\ .
\end{gather}
Combining these relations, we find
\begin{gather}
\sum_r \left(1-r\frac{q-2}{2 q}\right)f_r = q-1\ .
\end{gather}
Assuming that there are no snail diagrams, so that $f_1=0$, we have\footnote{Indeed, for any snail diagram, some of the double-line subgraphs must be non-planar. For $q=6$ this can be seen in fig. \ref{forget3color} by connecting a pair of fields and checking that some of the double-line propagators need to be twisted, thus causing non-planarity. For example, when connecting fields $1$ and $3$ the blue-green propagator clearly contains such a twist.}   
\begin{gather}
\frac{2}{q}f_2 = q-1 + \sum_{r>2} \left(r \frac{q-2}{2 q} - 1\right) f_r\ .
\end{gather}
For $q\geq 6$ the sum on the RHS of this equation is greater than zero. This implies that there is a loop passing through exactly two vertices. We shall call them a {\it basis} pair of vertices. Without a loss of generality one can assume that these vertices can be drawn as in fig. \ref{basispair}.  Also, for convenience we will number the fields in the vertices as in fig. \ref{basispair}. We can say that this loop, passing through two vertices, is a pair of bare propagators that connects the outputs with numbers $1_{L}$ with $1_{R}$ and $2_L$ with $2_R$, see fig. \ref{basispair}. Now let us choose any other field in the left vertex, $a_L$, in the range from $3_L$ to $q_L$ (for instance, we choose 3). Let us erase all colors except for $(1_L3_L)$ and $(3_L2_L)$. We can make a permutation of vertices such that the output will be between the first and second outputs (see fig. \ref{132fig}). However, the same does not hold for the right vertex; for example, between the $1_R$ and $2_R$ there could be another number of the field $r_i$, that must be non-zero.

\begin{figure}
	\centering
	{\includegraphics[width=0.35\textwidth]{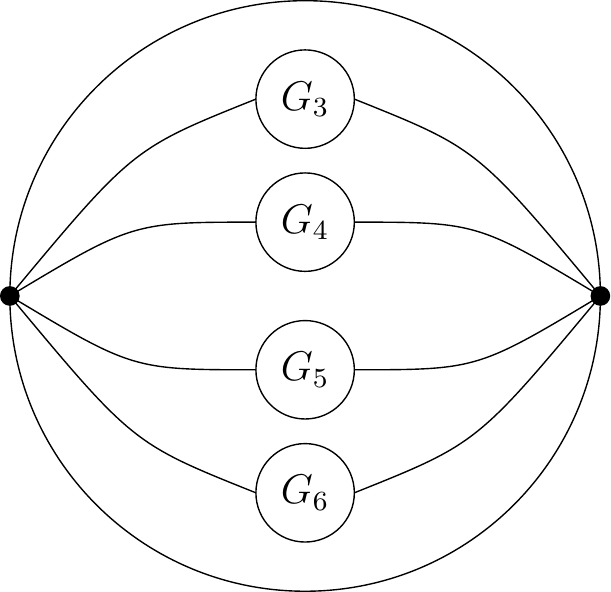}}
\caption{Any maximal graph for $q=6$ must be of this form. $G_i$ are arbitrary propagator insertions.}
\label{finalform}
\end{figure}

Because the double-line graph constructed out of the colors $(1_L 3_L)$ and $(3_L 2_L)$ should be planar, the output $3_L$ on the left vertex can be connected {\bf only} with these $r_i$ outputs. It cannot be connected with the other fields, and these $r_i$ fields in the right vertex could be connected {\bf only} to this field $3_L$ on the left (for example, in fig. \ref{132fig} the field $3_L$ can be connected only to the fields $3_R,5_R,4_R$ in order for the graph to be planar). From this we derive that for each field on the left we must assign a subset of the fields on the right. These subsets do not intersect with each other in order for the graph to be planar for any choice of the pairs of colors. From this we have
\begin{gather}
\sum_{a=3}^{q} r_a = q-2\ .
\end{gather}
Since $r_a \geq 1$, this equation implies $r_a=1$. Therefore, each output on the left is connected to the one on the right with a one-to-one correspondence. Thus, each ribbon graph, which is made by removing any set of $q-3$ colors, is planar. The graph has the structure depicted in fig. \ref{finalform} for $q=6$, where $G_i$ are propagator insertions.
We can connect the ends of these structures to get four other maximal vacuum diagrams and apply the same reasoning to them. From this one can see that the maximal graph must be melonic.  

Thus, we have shown that, in order for a graph to have the maximal large-$N$ scaling, it must be melonic.
It is also not hard to see \cite{Ferrari:2017jgw,Gubser:2018yec} that, if we take two MST interaction vertices and connect each field from one vertex with the corresponding field in the other, we will find the maximal large-$N$ scaling. This completes the argument that, for any MST interaction vertex, a graph has the maximal large-$N$ scaling if and only if it is melonic.

\section{Large-$N$ scaling dimensions of the fermion bilinears}\label{large_N}

Due to the melonic dominance for the rank $q-1$ tensor models with MST interactions, we can sum the Feynman diagrams in these large-$N$ theories. This allows us to calculate the propagator of fermionic fields and the spectrum of fermion bilinear operators. We expect the large-$N$ solution of the MST tensor models to be similar to that of the SYK models, which also exhibit the melonic dominance. 
Indeed, in \cite{Yoon:2017} it was shown that the four-point function for a rank $q-1$ tensor model has the same kernel as the SYK model four-point function with a $q$ fermion interaction. 
In this section we present further results along these lines.  

\begin{figure}
\centering
{\includegraphics[width=\textwidth]{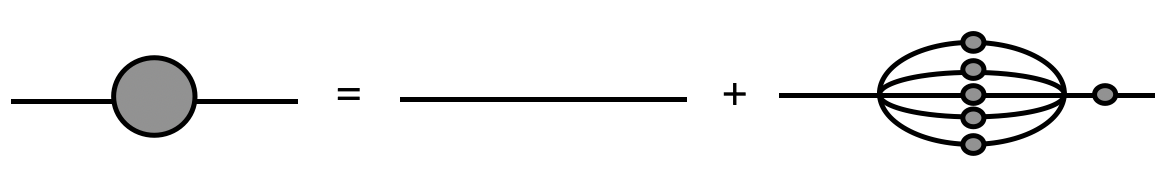}}
\caption{\label{fig:two_pt} Melonic corrections to the propagator. These are the only diagrams that survive at large $N$.}
\end{figure}

The large-$N$ Schwinger-Dyson equation for the tensor model two point function with a six fermion interaction is represented diagrammatically in fig. \ref{fig:two_pt}.
We can write the Schwinger-Dyson equations from the diagrams in fig. \ref{fig:two_pt}. We start with an MST $q$-tensor interaction,
\begin{equation}\label{DSequation}
G(t) = \braket{T\psi(t)\psi(0)} = \left(\partial_t - \Sigma\right)^{-1}, \quad\Sigma = q|\textit{Aut}| g^2 N^{\frac{(q-1)(q-2)}{2}} G^{q-1}, 
\end{equation}
where $\left|Aut\right|$ is the order of the automorphism group of the interaction (see section \ref{higher_q}), and $q|\textit{Aut}|$ is the number of contributing Feynman diagrams \cite{Gubser:2018yec}. We introduce $\lambda^2 = q|\textit{Aut}| g^2 N^{\frac{(q-1)(q-2)}{2}} $, and we make the assumption that in the IR regime the $\Sigma$ will dominate the derivative. Thus, we use the following conformal ansatz,
\begin{equation}
G(t) = \frac{a\, \sign(t)}{|t|^{2\Delta}}, \quad \Sigma(t) = \lambda^2 \frac{a^{q-1} \sign(t)}{|t|^{2(q-1) \Delta}}.
\label{anz}
\end{equation}
We take the Fourier transform of (\ref{anz}) and arrive at,
\begin{equation}
\begin{split}
&G(\omega) = 2^{1- 2 \Delta}i \sqrt{\pi} \frac{\Gamma(1-\Delta)}{\Gamma(1/2 + \Delta)} a |\omega|^{2\Delta-1} \sign(\omega),\\ \quad \Sigma(\omega) &= 2^{1- 2(q-1) \Delta}i \sqrt{\pi} \frac{\Gamma(1-(q-1)\Delta)}{\Gamma(1/2 + (q-1)\Delta)} a^{q-1} \sign(\omega) \lambda^2 |\omega|^{2(q-1) \Delta-1}.
\end{split}
\label{ft}
\end{equation}
In the IR limit we assume that we can neglect the derivative and get $G = -1/\Sigma$. From this we arrive at
\begin{gather}
-1 = G(\omega) \Sigma(\omega) = - 2^{2 - 2q \Delta}  a^q \lambda^2 \pi \frac{\Gamma(1-\Delta) \Gamma(1- (q-1)\Delta)}{\Gamma(1/2+\Delta) \Gamma(1/2 + (q-1) \Delta)} |\omega|^{2q\Delta-2}.
\end{gather}
It follows that $\Delta=1/q$ and $a^q  = \frac{\Gamma(\frac{1}{2}+\Delta)\Gamma(\frac{3}{2}-\Delta)}{\pi\lambda^2\Gamma(1-\Delta)\Gamma(\Delta)}$. Thus, we have that the propagator is,
\begin{gather}
G(t) = \left(\frac{\Gamma(\frac{1}{2}+\Delta)\Gamma(\frac{3}{2}-\Delta)}{\pi\lambda^2\Gamma(1-\Delta)\Gamma(\Delta)}\right)^\frac1q \frac{\sign (t)}{|t|^{2\Delta}},
\end{gather}
which exactly matches that of the general $q$ SYK model \cite{Gross2017}. For $q=6$ and $q=8$ we have,
\begin{gather}
G_6(t) = \left(\frac{\sqrt{3}}{9 \pi \lambda^2}\right)^\frac16 \frac{\sign(t)}{|t|^{1/3}}, \quad G_8(t) =\left(\frac{3}{8 \pi \lambda^2\text{cot}(\frac{\pi}{8})}\right)^\frac18 \frac{\sgn t}{|t|^{1/4}}.
\end{gather}
Using this propagator we can study the spectrum of bilinear operators. 

\begin{figure}
\centering
{\includegraphics[width=0.75\textwidth]{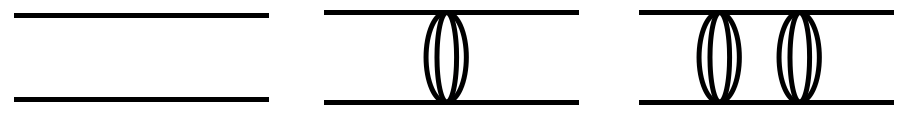}}
\caption{\label{fig:ladders} A few of the ladder diagrams that contribute to the four-point function.}
\end{figure}

Let us first compare the combinatorial factors in the ladder diagrams, shown in fig. \ref{fig:ladders}, to those in the melonic diagrams for the two-point function, shown in 
fig. \ref{fig:two_pt} . As stated above, there are $q|\textit{Aut}|$ Feynman diagrams that must be counted for each melon insertion. We note that the ladder diagrams may be constructed by `cutting' one of the internal legs of the melonic diagrams for the two-point function. There are $(q-1)$ choices of which leg to cut. This means that, for every diagram in fig. \ref{fig:two_pt}, we can make $(q-1)$ ladder diagrams by cutting the different internal propagators. So, we have a combinatoricial factor of $q(q-1)|\textit{Aut}|$ for the ladder diagram. Thus, the factors of $|\textit{Aut}|$ cancel in the operator spectra calculation, and we find that the spectrum is identical to that of the corresponding $q$ SYK model. The calculation is presented in the following.

The kernel comes from one rung of the ladder in fig. \ref{fig:ladders}. In the general $q$ case, we get that the kernel exactly matches that of the general $q$ SYK model \cite{Maldacena:2016hyu}:
\begin{gather}
\hat{K}_q=K_q(t_1,t_2;t_3,t_4) = - (q-1) \lambda^2 G(t_{13}) G^{q-2}(t_{34}) G(t_{24}). 
\end{gather}
For the $q= 6$ and $q=8$ case, we have
\begin{gather}
\begin{split}
\hat{K}_6&=K_6(t_1,t_2;t_3,t_4) = - 5 \lambda^2 G(t_{13}) G^4(t_{34}) G(t_{24}), \\ \quad
\hat{K}_8&=K(t_1,t_2;t_3,t_4) = - 7 \lambda^2 G(t_{13}) G^6(t_{34}) G(t_{24}).
\end{split}
\end{gather}

We substitute the ansatz for the spectrum of singlet bilinears as
\begin{gather*}
v_{i_1\ldots i_{q-1},j_1\ldots j_{q-1}}(t_1,t_2) = \braket{T O(\infty) \psi_{i_1\ldots i_{q-1}}(t_1)\psi_{j_1\ldots j_{q-1}}(t_2)} = \delta_{i_1 j_1}\ldots \delta_{i_{q-1}j_{q-1}}\frac{\sign (t_1 - t_2)}{|t_1 - t_2|^{2\Delta-h}},
\end{gather*}
where $h$ is the dimension of the operator $O(t)$.
The spectrum of operators for the $q = 6$ model is computed as follows: 
\begin{gather}
\hat{K}v(t_1,t_2)=\int dt_3 dt_4 K(t_1,t_2; t_3,t_4) v(t_3,t_4) =\notag\\= -\frac{5\sqrt{3}}{9\pi}\int dt_3 dt_4 \frac{\sign(t_1 - t_3)\sign(t_3 - t_4) \sign(t_4 - t_2) }{|t_1 -t_3|^{\frac13} |t_3 - t_4|^{\frac53-h} |t_4 - t_2|^{\frac13}} = g_a(h) v(t_1,t_2),\quad \text{where}\notag\\
g_a(h)= -5 \frac{\Gamma\left(\frac32-\Delta\right)\Gamma\left(1-\Delta\right)}{\Gamma\left(\frac12+\Delta\right)\Gamma\left(\Delta\right)}\frac{\Gamma\left(\Delta+\frac{h}{2}\right)\Gamma\left(\frac12+\Delta-\frac{h}{2}\right)}{\Gamma\left(\frac32-\Delta-\frac{h}{2}\right)\Gamma\left(1-\Delta+\frac{h}{2}\right)}\ ,
\end{gather}
and $\Delta=\frac16$.
The scaling dimensions of bilinear operators $\psi_{abcde}\partial_t^{2n+1}\psi_{abcde}$ are determined by the equation $g_a(h)=1$, 
and its form coincides with that for the SYK model \cite{Maldacena:2016hyu}:
\begin{equation}
\label{antisymgenq}
g_a(h)= -(q-1) \frac{\Gamma\left(\frac32-\Delta\right)\Gamma\left(1-\Delta\right)}{\Gamma\left(\frac12+\Delta\right)\Gamma\left(\Delta\right)}\frac{\Gamma\left(\Delta+\frac{h}{2}\right)\Gamma\left(\frac12+\Delta-\frac{h}{2}\right)}{\Gamma\left(\frac32-\Delta-\frac{h}{2}\right)\Gamma\left(1-\Delta+\frac{h}{2}\right)}\ , \qquad \Delta=\frac1q
\end{equation}
after setting $q=6$.
 There is a solution at $h=2$, which is the mode dual to the excitation in Jackiw-Teitelboim dilaton gravity \cite{Almheiri:2014cka,Jensen:2016pah,Maldacena:2016upp,Engelsoy:2016xyb}. One can show that the spectrum has the following asymptotic behavior, $h \to 2n+4/3$ as $n\to \infty$.
\begin{figure}
	\includegraphics[scale=0.85, trim=-6cm 0 0 -1cm]{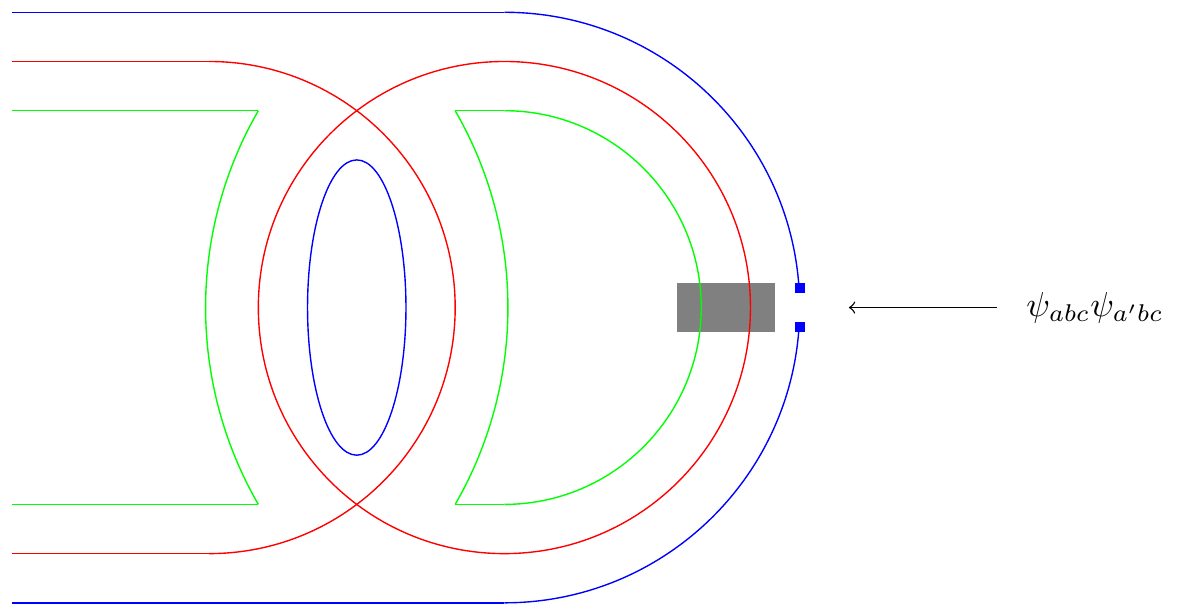}
	\caption{The insertion of the colored operator can suppress some diagrams, in contrast to the insertion of the singlet operator. For example, if one inserts the operator of the form $\psi_{ab \ldots yz}\partial_t^{2n+1}\psi_{ab\ldots yz'}$, only one diagram contributes in the large $N$ limit, compared to the $(q-1)$ contributions from a singlet operator.}
	\label{fig:LadCorCharge}
\end{figure}
In contrast to the SYK model, the tensor model contains operators which are $SO(N)$ symmetry generators, such as $J_{aa'} = \psi_{abcde}\psi_{a'bcde}$. If there are no ladder corrections to this operator, we would find that its scaling dimension is $\Delta_J = 2 \Delta_\psi = \frac13$; this would contradict the conservation of such charges. In fact, 
one can verify that there are ladder corrections to the operator which are non-vanishing in the melonic large $N$ limit \cite{Bulycheva:2017ilt} (see fig. \ref{fig:LadCorCharge}). 
Their feature is that, due to the antisymmetry in $a$ and $a'$, the relevant eigenfunctions are 
symmetric \cite{Bulycheva:2017ilt}:
\begin{gather}
v(t_1,t_2) = \braket{T O(\infty) \psi_I(t_1)\psi_J(t_2)} = \delta_{IJ}\frac{1}{|t_1 - t_2|^{1/3-h}}.
\end{gather}
Thus, we have
\begin{gather}
\hat{K}v(t_1,t_2)=\frac{-\sqrt{3}}{9\pi}\int dt_3 dt_4 \frac{\sign(t_1 - t_3) \sign(t_4 - t_2) }{|t_1 -t_3|^{\frac13} |t_3 - t_4|^{\frac53-h} |t_4 - t_2|^{\frac13}}=g_s(h)v(t_1,t_2).
\end{gather}
In general \cite{Bulycheva:2017uqj}, 
\begin{equation}
\label{symgenq}
g_s(h) = -\frac{\Gamma\left(\Delta-\frac{h}{2}\right)\Gamma\left(\Delta+\frac{h}{2}-\frac12\right)\Gamma\left(1-\Delta\right)\Gamma\left(3/2-\Delta\right)}{\Gamma\left(\frac12+\frac{h}{2}-\Delta\right)\Gamma\left(1-\Delta - \frac{h}{2}\right)\Gamma\left(\frac12+\Delta\right)\Gamma\left(\Delta\right)},\qquad \Delta=\frac1q\ , 
\end{equation}
and here we set $q=6$.

\begin{figure}
\centering
{\includegraphics[width=0.85\textwidth]{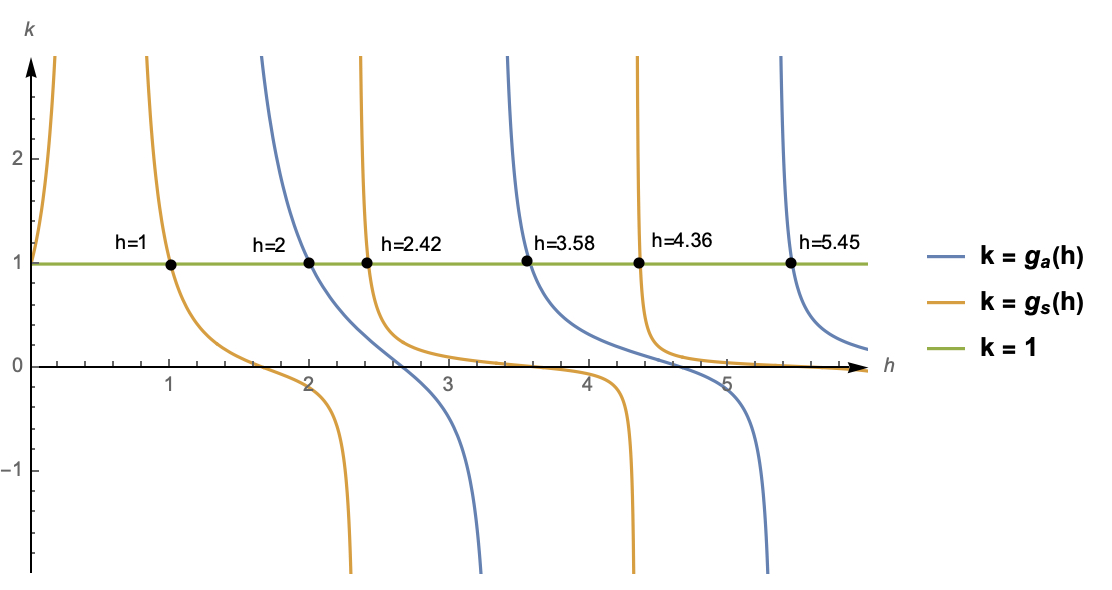}}
\caption{\label{fig:bilinears} The dimensions of bilinear operators in the $O(N)^5$ model.}
\end{figure}
 The equation for the scaling dimensions in the symmetric sector is $g_s (h) = 1$,
and one can check that $ h=0$ is a solution of this equation; it corresponds to a conserved charge. The asymptotic behavior of the eigenvalues is   $h \to 2n+1/3$, corresponding to
operators $\psi_{abcde}\partial_t^{2n}\psi_{a'bcde}$.

\begin{figure}
\centering\includegraphics[width=0.85\textwidth]{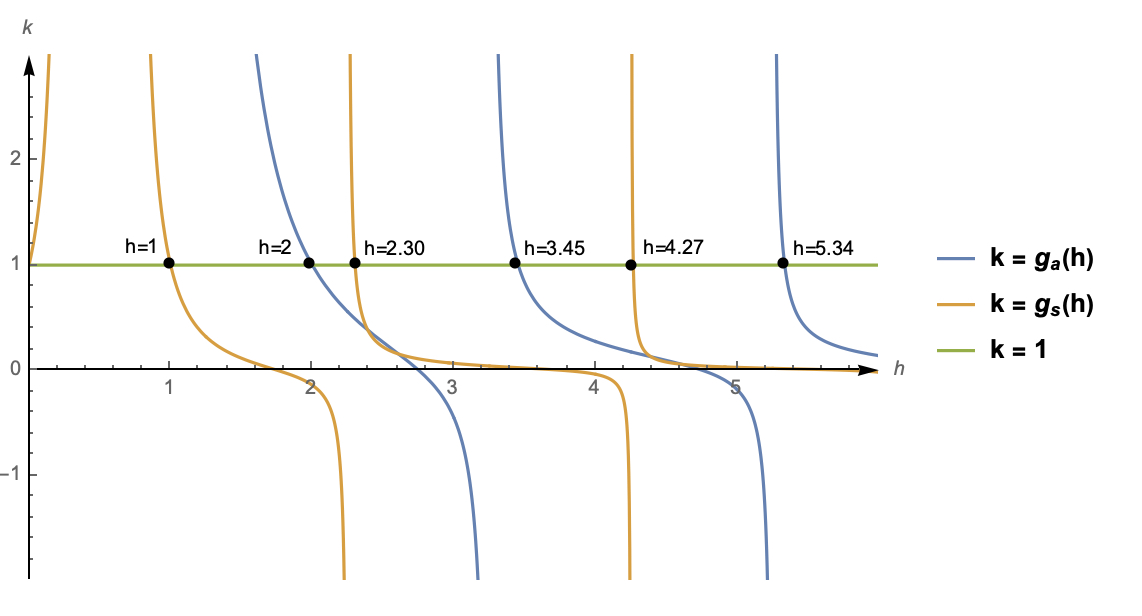}
\caption{\label{fig:gh}The dimensions of bilinear operators in the $O(N)^7$ model.}
\end{figure}

In an analogous manner, we can compute the spectrum of operators for $q = 8$,
\begin{gather}
\int dt_3 dt_4 K(t_1,t_2; t_3,t_4) v(t_3,t_4) = -\frac{21}{8 \pi \text{cot}(\frac{\pi}{8})}\int dt_3 dt_4 \frac{\sign(t_1 - t_3)\sign(t_3 - t_4) \sign(t_2 - t_4) }{|t_1 -t_3|^{\frac14} |t_3 - t_4|^{\frac74-h} |t_2 - t_4|^{\frac14}}  \notag\\
 = g_a(h) v(t_1,t_2)\ , 
\end{gather}
where $g_a(h)$ is given by (\ref{antisymgenq}) with $q=8$. The scaling dimension is determined by the equation $g_a(h)=1$. We can verify that there are no complex modes, that $g_a(h) = g_a(1-h)$, and that there exists a solution at $h = 2$, see fig. \ref{fig:gh}.

We can similarly examine the symmetric sector, where our ansatz is now,
\begin{gather}
v(t_1,t_2) = \frac{1}{|t_1 - t_2|^{1/4-h}}.
\end{gather}
Performing the analogous calculations, we find that,
\begin{gather}
\int dt_3 dt_4 K(t_1,t_2; t_3,t_4) v(t_3,t_4) = -\frac{3}{8 \pi \text{cot}(\frac{\pi}{8})}\int dt_3 dt_4 \frac{\sign(t_1 - t_3) \sign(t_2 - t_4) }{|t_1 -t_3|^{\frac14} |t_3 - t_4|^{\frac74-h} |t_2 - t_4|^{\frac14}}  \notag\\
 = g_s(h) v(t_1,t_2)\ ,
\end{gather}
and $ g_s(h) $ is obtained from (\ref{symgenq}) by setting $q=8$.

\section*{Acknowledgements}

This research was supported in part by the US NSF under Grant No.~PHY-1620059. 
We thank Nathan Benjamin, Raghu Mahajan, Vincent Rivasseau, Fidel Schaposnik Massolo, Grigory Tarnopolsky, Guillaume Valette, Edward Witten, and Junggi Yoon for useful discussions. 
We are especially grateful to Christian Jepsen for valuable discussions and comments on a draft of this paper.
IRK is grateful to the Kavli Institute for Theoretical Physics at UC, Santa Barbara and the organizers of the program ``Chaos and Order: From strongly correlated systems to black holes" for the hospitality and stimulating atmosphere during some of his work on this project. His research at KITP was supported in part by
the National Science Foundation under Grant No. NSF PH-1748958. The work of PNP is supported in part by the Dean's Grant from Princeton University. FKP thanks the organizers of the conference ``Quantum Gravity 2019" in Paris
for hospitality and useful discussions during the final stages of this project. 
 
\appendix

\section{$SO(N)^5$ invariant quartic operators } \label{quartic}
\begin{table}
	\centering
	\small
	\begin{tabular}{ | m{1.5cm} | m{.5cm} | m{.85cm} | m{1.2cm} | m{1.4cm} | m{1.7cm} | m{1.7cm} | m{2cm} | m{2.5cm} |}
		\hline
		operators & 1 & (1 2) & (1 2 3) & (1 2 3 4) & (1 2 3 4 5) & (1 2) (3 4) & (1 2) (3 4 5) & irreps\\ \hline
		\begin{minipage}{.3\textwidth}
			\vspace{1mm}
			\includegraphics[width=15mm, height=15mm]{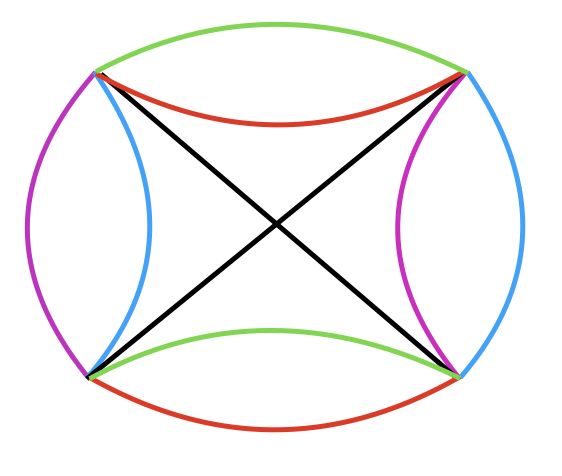}
			\vspace{1mm}
		\end{minipage}
		& 15 & 3 &  0 & -1 & 0 & -1 & 0 & \small{$\mathbf{4}\bigoplus \mathbf{6}\bigoplus \mathbf{5}$}
		\\
		\hline
		\begin{minipage}{.3\textwidth}
			\vspace{1mm}
			\includegraphics[width=15mm, height=15mm]{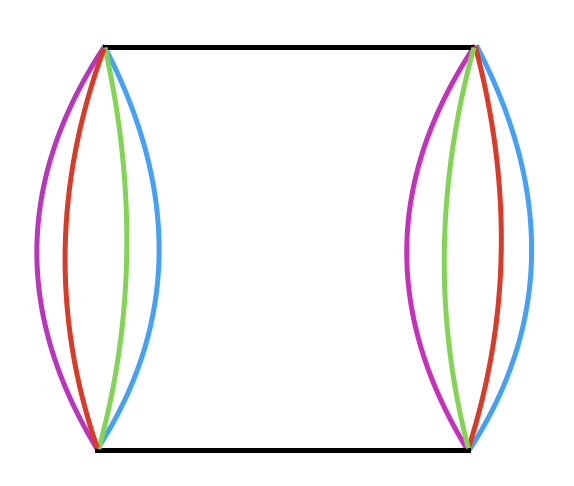}
			\vspace{1mm}
		\end{minipage}
		& 5 & 3 & 2 & 1 & 0 & 1 & 0 & \small{$\mathbf{1}\bigoplus \mathbf{4}$}
		\\ \hline
		\begin{minipage}{.3\textwidth}
			\vspace{1mm}
			\includegraphics[width=15mm, height=15mm]{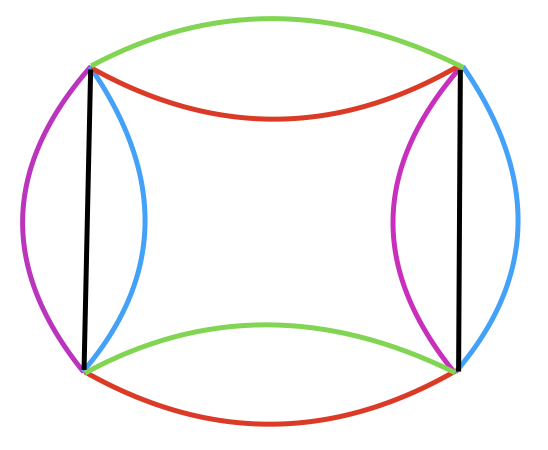}
			\vspace{1mm}
		\end{minipage}
		& 10 & 4 & 1 & 0 & 0 & 2 & 1 & \small{$\mathbf{1}\bigoplus \mathbf{4}\bigoplus \mathbf{5}$}
		\\ \hline
		\begin{minipage}{.3\textwidth}
			\vspace{1mm}
			\includegraphics[width=15mm, height=15mm]{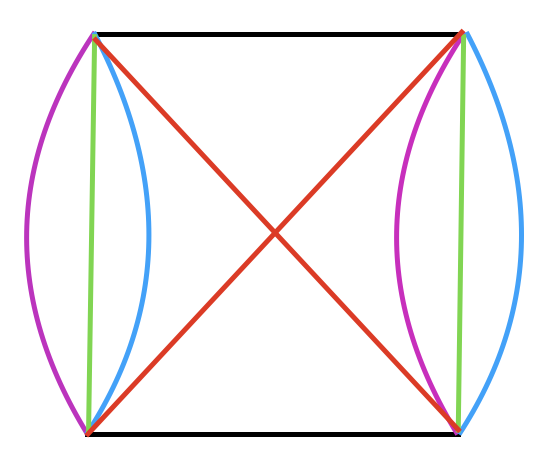}
			\vspace{1mm}
		\end{minipage}
		& 10 & 2 & 1 & 0 & 0 & -2 & -1 & \small{$\mathbf{4}\bigoplus \mathbf{6}$}
		\\ \hline
	\end{tabular}
	\caption{Character Table for Quartic Operators. $\mathbf{1}$ is the trivial representation, $\mathbf{4}$ is the standard representation, $ \mathbf{6}$ is the exterior square of the standard representation, and $\mathbf{5}$ is the irreducible 5 dimensional representation.}
	\label{tbl:char}
\end{table}
In this appendix we classify the $SO(N)^5$ invariant quartic operators in the theory  (\ref{Hamil})
according to their transformational properties under the action of the discrete symmetry $S_5$ discussed in section \ref{sub_discrete}. 
We will show that these operators do not transform nicely under the $A_5\subset S_5$ symmetry which consists of the even permutations of the five $O(N)$ groups. In order to find the possible singlet quartic operators, we must find all the distinct ways the indices of the four fermions may be contracted. We pictorially represent the quartic operators of the theory in table \ref{tbl:char}. We represent each fermion as a vertex and the index contractions are represented by edges connecting the distinct vertices. We can denote the number of edges connecting each vertex to the others by three integers  $\rho_a$, $\rho_b$, and $\rho_c$. To find the possible  quartic operators  in this theory we consider all possible combinations of integers $\rho_a$, $\rho_b$, and $\rho_c$ that satisfy the relations that the number of edges at each vertex is five ($\rho_a+\rho_b+\rho_c = 5$) and that a fully connected quartic operator must not have more than four strands shared between two nodes ($5>\rho_a\geq \rho_b\geq \rho_c\geq 0$). We find the following triples: $(4,1,0)$, $(3,2,0)$, $(3,1,1)$, and $(2,2,1)$. Each triplet corresponds to the construction of a quartic term displayed in table \ref{tbl:char}, and they are of the following form: $\psi^{a_1b_1c_1d_1e_1}\psi^{a_2b_2c_1d_1e_2}\psi^{a_1b_1c_2d_2e_2}\psi^{a_2b_2c_2d_2e_1}$ corresponding to $(2, 2, 1)$ in row 1, $\psi^{a_1b_1c_1d_1e_1}\psi^{a_2b_2c_2d_2e_1}\psi^{a_1b_1c_1d_1e_2}\psi^{a_2b_2c_2d_2e_2}$ corresponding to $(4, 1, 0)$ in row 2, $\psi^{a_1b_1c_1d_1e_1}\psi^{a_2b_2c_2d_1e_1}\psi^{a_1b_1c_1d_2e_2}\psi^{a_2b_2c_2d_2e_2}$ corresponding to $(3,2,0)$ in row 3, and $\psi^{a_1b_1c_1d_1e_1}\psi^{a_2b_2c_2d_1e_2}\psi^{a_1b_1c_1d_2e_2}\psi^{a_2b_2c_2d_2e_1}$ corresponding to $(3, 1, 1)$ in row 4. 

Now we can find the irreducible representations of $S_5$ of each of the possible quartic operators and show that none transform nicely under $A_5$. We use character theory to do this. We must consider the number of fixed points (the character) of each of the operators under the action of the conjugacy classes of $S_5$, which are included in the top row of table \ref{tbl:char}.  The negative values represent the exchange of an odd number of vertices of the operator under the conjugacy class. By calculating the inner products of the characters of the operators with the characters of the irreducible representation, we can find the correct group decomposition \cite{fulton}.
The possible quartic operators of $O(N)^5$, their character tables, and irreducible representations of $S_5$ are summarized in table \ref{tbl:char}.

\bibliographystyle{ssg}
\bibliography{sixorder}

\end{document}